\documentstyle{mn}
\input psbox.tex

\def\go{
\mathrel{\raise.3ex\hbox{$>$}\mkern-14mu\lower0.6ex\hbox{$\sim$}}
}
\def\lo{
\mathrel{\raise.3ex\hbox{$<$}\mkern-14mu\lower0.6ex\hbox{$\sim$}}
}

\def\eg{{\it e.g.\ }}
\def\ie{{\it i.e.\ }}
\def\vinf{V_\infty}
\def\rmin{R_{\rm min}}

\def\deltaee{\Delta E/E\,}
\def\etal{{\it et al.\ }}

\begin{document}

\title{Red-giant collisions in the galactic centre}

\author[Bailey and Davies]{Vernon C. Bailey, Melvyn
B. Davies\thanks{Current 
Address: Department of Physics and Astronomy,
 University of Leicester, Leicester LE1 7RH}\\
Institute of Astronomy, Madingley Road, Cambridge CB3 OHA.}

\date{Received ** *** 1998; in original form 1998 *** **}

\maketitle

\begin{abstract}

We simulate collisions involving red-giant stars in the centre of our
galaxy. Such encounters may explain the observed paucity of highly
luminous red giants within $\sim 0.2\,$pc of Sgr A$^\star$. The masses of
the missing stars are likely to be in the range $\sim 2\,$M$_\odot\,-\,8
\,$M$_\odot$. 
Recent models of the galactic centre cluster's 
density distribution and velocity dispersion
are used to calculate two-body collision rates. 
In particular we use stellar-evolution models to calculate
the number of collisions a star will have during different 
evolutionary phases. We find that the number of two-body 
collisions per star is $\lo 1$ in the central $0.1\,$pc$\,-\,0.2\,$pc, 
depending strongly on the galactocentric radius, with some
uncertainty from the assumed cluster models and stellar-evolution
models. Using a 3D numerical hydrodynamics code (SPH) we simulate
encounters involving cluster stars of various masses with a $2\,$M$_\odot$
red giant and an $8\,$M$_\odot$ red giant. The instantaneous mass loss in such
collisions is rarely enough to destroy either giant. A fraction of
the collisions do, however, lead to the formation of common envelope
systems where the impactor and giant's core are enshrouded by the
envelope of the giant. Such
systems may evolve to expel the envelope, leaving a tight binary; the
original giant is destroyed. 
The fraction of collisions that produce common envelope
systems is sensitive to the local velocity dispersion and hence
galactocentric radius. Whereas most of our collisions lead to common envelope
formation at a few parsecs from Sgr A$^\star$, very few collisions do
so within the central $0.2\,$pc. Using our collision-rate calculations we then compute
the time-scales for a giant star to suffer such a collision within the
galactic centre. These time-scales are $\go 10^{9-10}$ years and so
are longer than the lifetimes of stars more-massive than $\sim
2\,$M$_\odot$. Thus the observed paucity of luminous giants is unlikely to be
due to the formation of common envelope systems as a result of
two-body encounters involving giant stars. 
\end{abstract}


\section{Introduction}

Recent observations of the central stellar cluster of our
galaxy reveal that the brightest K-giant stars
($2\,-\,8\,$M$_\odot$ AGB stars or more-massive supergiants) are depleted
within $\sim 0.2\,$pc from Sgr A$^{\star}$, the dynamical centre of the
Milky Way (Genzel \etal 1996).  The missing stars are highly luminous
and are therefore very large, and so 
Genzel \etal (1996) concluded that 
stellar collisions were a likely explanation for the paucity if the cluster's
core density is sufficiently large to make collisions frequent
and if such collisions permanently destroyed the photospheres of
giants. 

The effects of stellar collisions on giant stars have been previously
investigated in the context of globular clusters (Davies, Benz \&
Hills 1991; 1992). Such collisions often resulted in the impactor
becoming bound to the giant's core within the envelope of the giant,
forming a {\it common envelope} system. In such systems, the impactor
and the core spiral 
towards each other and the binding energy released is deposited in the
envelope which is ejected on time-scales shorter than the evolutionary
time-scale of the giant. By this method the giant is destroyed,
leaving behind a binary comprised of the impactor and the giant's
core. Collisions in galactic centres differ from those in globular
clusters due to the higher velocities involved. 
The higher velocity dispersions in galactic nuclei
($\sigma\go 100\,$km/s) mean that incoming stars are faster and so are less
likely to form bound systems than in globular cluster cores
($\sigma\sim 10\,$km/s).  

In this paper we investigate the frequency of two-body collisions involving
main-sequence stars and also red giants. The red giants investigated
are similar to those observed to be depleted in 
the galactic centre (hereafter GC). 
We simulate a number of encounters involving the giants,
quantifying the mass loss and probability of forming bound systems. 
If the GC contains
binary stars, the collision time-scale between giants and binaries may
be significant if the cluster density is
sufficiently large. In a companion paper,
Davies \etal (1998), we extend this study
to collisions involving binary impactors.

In $\S$2 we consider the stellar contents, motions and distribution
within the central parsec of our galaxy, and we 
discuss the possible nature of the
stars that currently cannot be detected. The stellar density profile
and velocity dispersion are then used to calculate collision
time-scales in $\S$3. In $\S$4 we consider how devastating an encounter
must be to destroy a giant. We discuss
our approach to simulating collisions involving red giants in $\S$5 and
we present our results in $\S$6. In a number of collisions the
impactor will become bound to the giant star. We comment on the nature of
such systems in $\S$7, prior to
discussing our results in $\S$8. Alternative methods of destruction
for the giants are explored in $\S$9. We summarise our findings in
$\S$10.

\section{Stellar contents, distribution \& dynamics of the central parsec}

\subsection{Observable stars and the bright-giant paucity}

Dust in the plane of our galaxy obscures the GC by $\sim 30$
magnitudes of visual extinction (Rieke, Rieke \& Paul 1989). The
infra-red K band ($2.2\,\mu$m) extinction is much less severe,
A$_{\rm K}\sim3$ magnitudes (Rieke, Rieke \& Paul 1989). The large distance
to the GC ($\sim 8.5\,$kpc) gives a distance modulus of 14.6
which, coupled with A$_{\rm K}\sim3$, limits stellar observations to giant
stars. Within the central $\sim 0.5\,$pc of our galaxy there are
$\sim 600$ red giants of magnitude K$\,\lo 15$ (Eckart \etal
1995). However, the brightest red giants are {\it not observed} in the
deepest parts of the galactic nucleus. Observations show sharp drops
in the numbers of $10.5<\,$K$\,<12$ stars at $\sim 0.12\,$pc and of
K$\,< 10.5$ stars at
$\sim 0.2\,$pc (Genzel \etal 1996).  
These highly luminous stars are unlikely to have been overlooked by
observations. Indeed, the number counts in
Genzel \etal (1996) of $10.5<\,$K$\,<12$ giants are likely to be 
$80\,-\,85\%$ complete 
and the numbers of the K$\,< 10.5$ stars are $90\,-\,95\%$
complete (Genzel, private communication). 
At slightly larger galactocentric radii, giants of similar
luminosity to those missing have a surface-density profile, 
$\Sigma \propto p^{-1}$, for projected radius $p$. Assuming that this profile
should extend further into the core, we estimate the numbers of
missing giants to be twenty-five K$\,< 10.5$ magnitude stars and 
eleven $10.5<\,$K$\,<12$ magnitude stars. If the surface
density were flat within the region of missing giants then lower
limits of the numbers of missing K$\,< 10.5$ and $10.5<\,$K$\,<12$ giants are
eleven and six respectively. 
We note that the {\it precipitous} drop, rather than
a gradual one, indicates that some mechanism has actively destroyed at
least $\sim 15\,-\,40$ stars in the central $0.2\,$pc.

\subsection{The visible stellar population}

\begin{figure}
\psboxto(\hsize;0cm){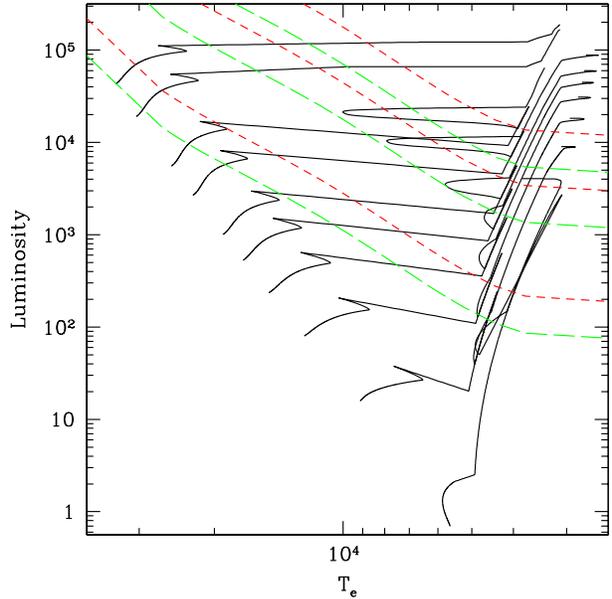}
\caption{HR diagram showing theoretical evolutionary tracks for stars
of masses 1, 2, 3, 4, 6, 8, 10, 15, 20$\,$M$_\odot$, models based upon
Pols \etal (1999) with metallicity Z=0.02. The full models include
evolution from AGB tip to remnant; we have truncated the tracks at the
AGB tip on this diagram for clarity. 
 The long-dashed lines represent contours of K of 15,
12 and 10.5, based on A$_{\rm K}=3$ and a distance modulus of 14.6 to the
GC; the short-dashed lines are the K
contours of the same values but based upon A$_{\rm K}=4$.  Effective temperature is in
Kelvin, luminosity is solar.}
\label{fig:HR1}
\end{figure}

\begin{figure}
\psboxto(\hsize;0cm){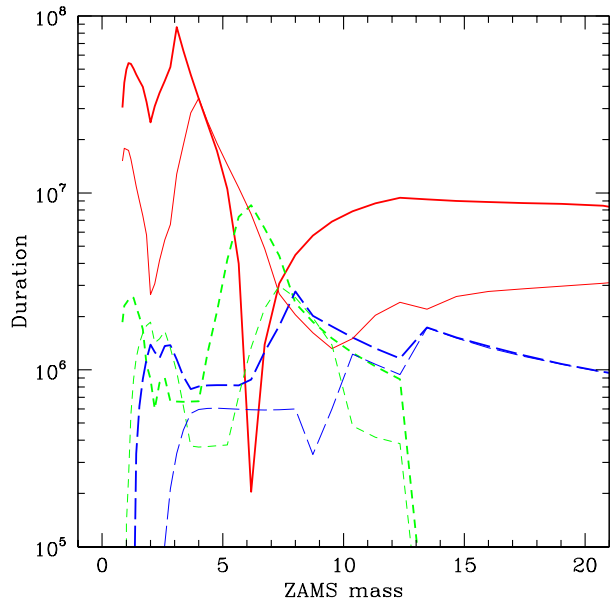}
\caption{Visible time-scales for stars at the GC, based on stellar
models as described in Figure~\protect\ref{fig:HR1}. The solid line denotes
the time-scale
for which the star is visible as $15<\,$K$\,<12$, the short-dashed line
represents the time-scale for which the star is visible as
$12<\,$K$\,<10.5$ and the long-dashed line shows the visible duration 
in the K$\,< 10.5$ band. Heavy lines show the time-scales for A$_{\rm
K}=3$; less bold lines represent A$_{\rm K}=4$. The
``kinks'' in the profiles can be attributed to different parts of
the stars' evolution (\eg horizontal branch) overlapping with a
given K-luminosity band. Note that
stars of masses less than $\sim 0.9\,$M$_\odot$ will not ascend the giant
branch within a Hubble time.}
\label{fig:times}
\end{figure}

The outcome of a collision between two stars will depend on their
masses and sizes (\ie evolutionary phases). It is useful to visualise
the types of stars detectable (and those which are missing) at the GC using
a HR diagram. For a star of given temperature and bolometric
luminosity, its apparent K-magnitude may be calculated using
bolometric corrections and V-K colours (Johnson 1966) and by assuming
a distance modulus and a K-extinction law. Taking the galactocentric
distance to be $8.5\,$kpc, Figure~\ref{fig:HR1} shows a HR digram with
contours of constant K appropriate for the GC, assuming  A$_{\rm K}=3$
and A$_{\rm K}=4$. The effects of uncertainties in the GC distance on
the masses and numbers of stars one expects to see at the GC are small
compared with uncertainties in A$_{\rm K}$. Rieke,
Rieke \& Paul (1989) report A$_{\rm K}=2.7$; more recently Blum,
Sellgren \& DePoy (1996) have derived a mean K-extinction of
$3.3\pm 0.9$ but note that A$_{\rm K}$ may reach six in places. 
We see from Figure~\ref{fig:HR1} that, for both values of A$_{\rm K}$, the
cool, visible stars must be giants and that we cannot easily determine
their masses because of the closeness of the Hayashi tracks. The
Figure does, however, imply that a visible low-mass star will likely be on
the AGB phase, whilst for
stars of increasing masses the horizontal branch becomes
increasingly visible. The duration for which a star of a
given K-magnitude is visible is an irregular function of its mass
(Figure~\ref{fig:times}) depending on whether either the long-lived 
main-sequence or
horizontal-branch phases coincide with the
chosen K-magnitude. Figure~\ref{fig:times} highlights the great
sensitivity of the visible durations of stars on 
A$_{\rm K}$. If A$_{\rm K}=3$, stars of masses $\lo 1.4\,$M$_\odot$ cannot
be detected at K$\,<10.5$. If A$_{\rm K}=4$, 
then stars of mass $\lo 2.5\,$M$_\odot$ cannot be detected
at K$\,< 10.5$ and the time-scale for which the stars are visible falls
sharply for masses $< 4\,$M$_\odot$.
Assuming A$_{\rm K}=3$, stars of most masses are brighter than K=10.5 for about
$10^6$ years and most stars of magnitude
$10.5<\,$K$\,<12$ are visible for
$6\,-\,85\,\times\,10^5$ years. For stars of masses in the region of
$6\,$M$_\odot$, the horizontal branch shines in the $10.5<\,$K$\,<12$ band
rather than in the $12<\,$K$\,<15$ band. We thus see a steep drop in the
time that such stars spend in the fainter band, and a
corresponding peak in the brighter band. The $10.5<\,$K$\,<12$ range corresponds
to the Hertzsprung Gap for stars of higher masses and so the
$10.5<\,$K$\,<12$ durations drop.
 The bulk of detectable stars sit in the $12<\,$K$<\,15$ bin
in which low-mass stars are visible for $\sim 3\,-6\,\times\,10^7$ years (still
assuming that A$_{\rm K}=3$); higher-mass stars are visible for
$10^7$ years. The fluctuation in visible duration for low-mass stars 
in the $12<\,$K$<\,15$ range is due to changes in positioning of the
core-helium burning phase with respect to the K$\,=15$ contour.
Increasing A$_{\rm K}$ to four generally decreases the visible
time-scales by a factor of a few (but within an order of
magnitude). Some stars may have their visible time-scales boosted in
some K-magnitude range because, say, their horizontal branches now
correspond to a lower K-magnitude bin in which, for lower A$_{\rm
K}$, only part
of their giant branches had been detectable. 

\subsection{The entire stellar population}

The contours of constant K-magnitude on Figure~\ref{fig:HR1} show that
detectable stars must be either evolved stars of a range of masses or
very massive stars if they are on the main sequence. If the GC
population consists mainly of low-mass main-sequence stars and stellar
remnants then we only glimpse a small fraction of the total mass of
the GC stellar
population. Indeed, the detectable stars' motions and distribution
have been used to measure the total mass in the central parsec to be
$\sim 4.2\,\times\,10^6\,$M$_\odot$, of which
$2.5\,\times\,10^6\,$M$_\odot$ is the
mass of the putative supermassive black hole at the centre of
the cluster (Genzel \etal 1996, Eckart \& Genzel 1997). The amount of
stellar mass in the central parsec is therefore about
$1.7\,\times\,10^6\,$M$_\odot$. We saw in $\S$2.2 that only stars of masses
$\go 1.4\,$M$_\odot$ may be seen in the GC if A$_{\rm
K}=3$. Assuming a Scalo IMF, stars of masses $>1.4\,$M$_\odot$ can account
for about a quarter of the mass of the population, if the mass contribution
from stellar remnants is small. Figure~\ref{fig:times} shows that
low-mass giants may be detectable for about $4\,\times\,10^7$ years,
depending on the stellar mass. Given that a $1.4\,$M$_\odot$ star lives
for $\sim\,4\,\times\,10^9$ years (Pols \etal 1999), for a constant
star-formation rate one might expect to see $\sim\,3000$ stars within
the central parsec. This is about a factor of two higher than, but is
not wildly inconsistent with, the
number counts one would observe by extrapolating the observed numbers
of stars within $0.5\,$pc (Eckart \etal 1995) out to a radius of $1\,$pc. 

A more detailed population synthesis was performed by Genzel \etal
(1994); they found that the GC stellar contents were inconsistent with
a constant star-formation history and Salpeter IMF. Detailed analyses
are, however, hampered by uncertainties in A$_{\rm K}$, possible
incompleteness in the observed number counts and that there may
not be a unique solution of IMF or star-formation history (or of both)
which reproduces the observations. 

The situation is even further
complicated by the possibility of a raised low-mass cut-off to star
formation at the GC (Morris 1993). For a cut-off of $\sim\,0.8\,$M$_\odot$ and
a constant star-formation history, one might expect roughly equal
numbers of white dwarfs and
main-sequence stars. Higher low-mass cut-offs may allow stellar
remnants to dominate the population, assuming that the population is
old. A further complication is that the GC stellar
contents may not be totally indigenous. Morris (1993) calculates the
mass fractions of neutron stars and black holes that may have
segregated into the central parsec from outside. 
Solutions computed by Morris (1993) allow most of the GC mass
to be accounted for by black holes and neutron stars. 
These calculations, however, suffer from requiring some parameters
that are not well constrained,
such as the low-mass cut-off and the mass of stellar black
holes. For example, lowering the mass of stellar black
holes from $10\,$M$_\odot$ to $3\,$M$_\odot$ and adopting a low-mass cut-off of
$0.08\,$M$_\odot$ reduces the mass-fraction of neutron stars and
stellar-mass black holes to $<10\%$ by mass. There will almost
certainly be non-indigenous remnants in the GC, but in quantities which
are currently poorly constrained.

The overall mass function of the GC population is therefore very
uncertain. For A$_{\rm K}=3$, we might expect most of
the missing giants to be of low masses (say,
$1.4\,$M$_\odot\,-\,3\,$M$_\odot$),
 for most
IMFs. However, the uncertainty in A$_{\rm K}$ could mean that the
brightest missing stars are of masses $\go 4\,$M$_\odot$.

\subsection{Distribution and motions of GC stars}

An observed number density of stars traces that of the whole cluster
if the mean mass of the cluster stars is close to the mean mass of the
observed stars and if the system is relaxed. The relaxation time 
(see Binney \& Tremaine 1987) at $0.2\,$pc is
$\sim\,2\,\times\,10^9$ years for a $1\,$M$_\odot$ star, so we expect low-mass
($1\,-2\,$M$_\odot$) giants to trace the overall distribution of stars of
similar  masses. Relaxation time decreases with the mass of the star but
more-massive stars do not live long enough to become relaxed in the
GC. Such stars
should be treated cautiously as mass tracers. Genzel \etal (1996)
model the observed red-giant number-density distribution to be 
\begin{equation}
n(r) \propto \biggl({1 \over 1 + (r/4.2^{''})^{1.8}}
\biggr)
\label{eqn:ndens}
\end{equation}
for galactocentric radius, $r$.
We adopt this Equation, assuming that the observed stars are indeed
tracing the whole population. This assumption is somewhat supported by
the stars' isotropic motions (Genzel \etal 1996, Eckart \& Genzel
1997), indicating a relaxed population. Genzel \etal (1996) also give
the central stellar density, $\rho_0$, to be
$4\,\times\,10^6\,$M$_\odot/$pc$^{3}$ and a core radius of $0.38\,$pc.  We
normalise Equation~(\ref{eqn:ndens}) with this value of $\rho_0$ but
note that $\rho_0$  and the core radius are interlinked values and
that the latter is subject to some debate, depending on whether the
cluster's surface-brightness distribution or the surface-density
distribution is used (see Genzel \etal 1994). By varying $\rho_0$ and
the core radius between $0.25\,$pc and $0.5\,$pc whilst conserving the total
central-parsec mass within errors set by Genzel \etal (1996), we find
that the collision time-scales for stars (see $\S$3) can vary by up to
a factor of 1.5.  

We also note that Equation~(\ref{eqn:ndens}) gives quite a flat profile
inside the core radius. However, observations become crowded within
this radius and it becomes harder to constrain the density
distribution. If the condensed mass at Sgr A$^\star$ is a black hole,
theoretical modelling predicts a density cusp at low radii: $\rho
\propto r^{-\gamma}$ where $\gamma\sim\,1.5\,-\,2.5$ (Quinlan, Hernquist \&
Sigurdsson 1995; see also Bahcall \& Wolfe 1976).  Consequently we
also calculate collision rates assuming $\rho \propto r^{-1.8}$
(coinciding with the higher-radius density profile at the GC), 
which we set equal to the density profile of Genzel \etal (1996) for $r\geq
0.38\,$pc where we expect the latter to be accurate.

A final comment on Equation~(\ref{eqn:ndens}) is that it is based upon
(M/L)$_{\rm K}=2$, the mass-to-light ratio evaluated at a few parsecs
from the centre. Genzel \etal (1996) claim that their usage of only brighter
cluster members reduces the effects of population change at lower
radii. There does exist some evidence of an increase in (M/L)$_{\rm K}$ 
at low radii: Saha \etal (1996) show that this ratio may
increase by a factor of 3-5 between $2.5\,$pc and $0.5\,$pc, although if
one subtracts $2.5\,\times\,10^6\,$M$_\odot$ for the putative central black
hole then the increase in (M/L)$_{\rm K}$ {\it for the cluster only} reduces
to two.

Collision-rate calculations require a velocity dispersion (see
$\S$3). We adopt the deprojected velocity dispersion given by
Genzel \etal (1996):
\begin{equation}
\sigma^{\rm 2} = 55^{\rm 2}+103^{\rm 2}(r/10'')^{\rm -1.2} \; {\rm (km/s)^{2}}.
\label{eqn:sig1}
\end{equation}
The velocity dispersion rises as galactocentric radius falls due to
the presence of the supermassive black hole (see also Eckart \& Genzel
1997). At higher radii, the
enclosed cluster mass dominates and $\sigma$ flattens
out. Equation~(\ref{eqn:sig1}) is valid throughout the central parsec.

\section{Collision rates of the central-parsec stars}

The collision rate per star of species 1 with stars of species 2 is
\begin{equation}
\Gamma_1={\Gamma_{12}(r)\over n_1(r)} = n_2(r) \int_{0}^{\infty} 
\Sigma_{\rm g}(V_{\infty}, R_{\rm min}) V_{\infty}
P(V_{\infty})\, dV_{\infty}
\label{eqn:gam}
\end{equation}
where $n$ is the number density of stars and $P(V_\infty)$ is the
probability that two stars have {\it relative} speed
$V_\infty$. $\Sigma_{\rm g}$ is the gravitationally focused cross
section:
\begin{equation}
\Sigma_{\rm g} =\pi R_{\rm min}^{\rm 2}\biggl( 1 +
{2G(M_{\rm 1}+M_{\rm 2})\over R_{\rm min}V_{\infty}^{\rm 2}}\biggr)
\label{eqn:gsig}
\end{equation}
where $M_1$ and $M_2$ are the stellar masses and $R_{\rm min}$ 
is the collision radius (which we
take to be equal to the sum of the radii of the two colliding stars). 
The second term on
the right in $\Sigma_{\rm g}$ 
represents the mutual attraction between two stars
which increases the cross section above that of the geometric
value. We take $P(V_\infty)$ to be Maxwellian, which is appropriate
at high galactocentric radii where the stellar cluster dominates
the enclosed mass. In the region where the black hole dominates the
enclosed mass (say, $\lo 0.7\,$pc, see Genzel \etal 1996, Eckart \& Genzel
1997), $P(V_\infty)$ is still close to Maxwellian (Quinlan \etal
1995). At any galactocentric radius we take the {\it relative} velocity
dispersion between two stellar species of any mass to be $2\sigma^2$,
where the variation of $\sigma$ with galactocentric radius is given by
Equation~(\ref{eqn:sig1}). 
In a system of stars that has achieved equipartition of energy, we have 
$\sigma^2\propto 1/M$ (see Binney \& Tremaine 1987); stars of
different masses would have different velocity dispersions. The
effects of equipartition on the collision rates do not change our
conclusions.

\begin{figure}
\psboxto(\hsize;0cm){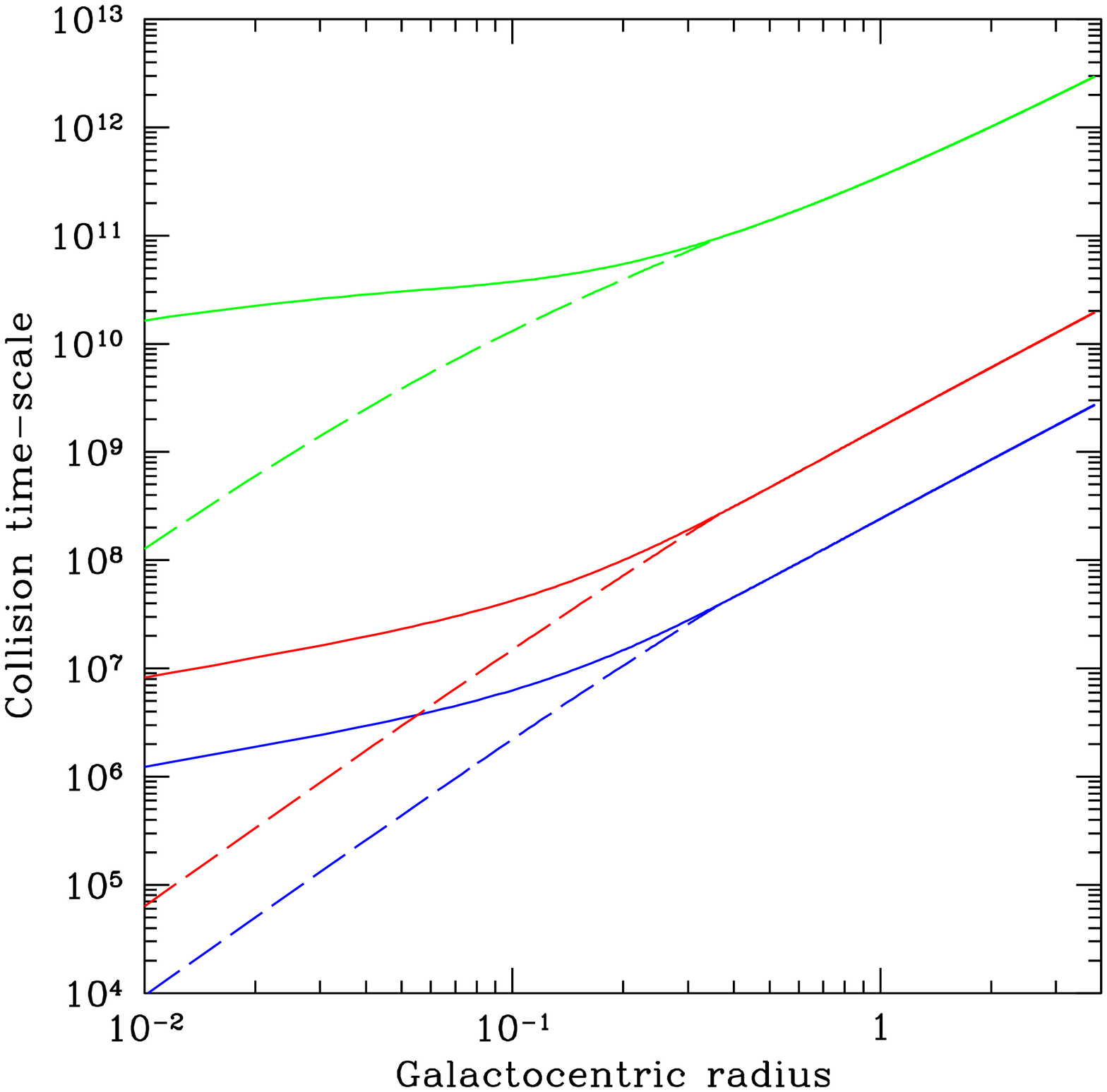}
\caption{Collision time-scales (in years) for stars in the GC
as a function of galactocentric radius (in parsecs). For
the calculation we have assumed that the background cluster members
are of mass $1\,$M$_\odot$, Equation~(\protect\ref{eqn:sig1}) for the velocity
dispersion, and for the density profile we have used
Equation~(\protect\ref{eqn:ndens}) (solid lines) or $\rho\propto r^{\rm -1.8}$
(dashed lines), set equal to Equation~(\protect\ref{eqn:ndens}) for $r\geq
0.38\,$pc. The
two lower  lines represent the time-scale for encounters for 
$8\,$M$_\odot$ stars at $R_{\rm min}=243\,$R$_\odot$. The middle two
curves represent  the time-scale between collisions for $2\,$M$_\odot$
stars at $R_{\rm min}=94\,$R$_\odot$. 
For comparison, the upper-most
lines show the
encounter time-scale profile for a $1\,$M$_\odot$ star at $R_{\rm
min}=2\,$R$_\odot$.}
\label{fig:timesc}
\end{figure}

Assuming number-density and velocity-dispersion profiles discussed in
$\S$2, Figure~\ref{fig:timesc} shows collision time-scales for the GC.
The collision time-scale is a strong function of $r$ and is shortest 
deep within the core. 
Comparing Figure~\ref{fig:timesc} with Figure~\ref{fig:times} we
see that the time-scales for collisions involving giants are similar to
the durations for which they are visible at the GC. The collision
time-scale for solar-type stars is considerably longer. It is only
noticeably less than a Hubble time at very low radii if the density
profile is a cusp (\ie within $\sim\,0.1\,$pc, assuming $\rho \propto
r^{-1.8}$).  
\begin{figure}
\psboxto(\hsize;0cm){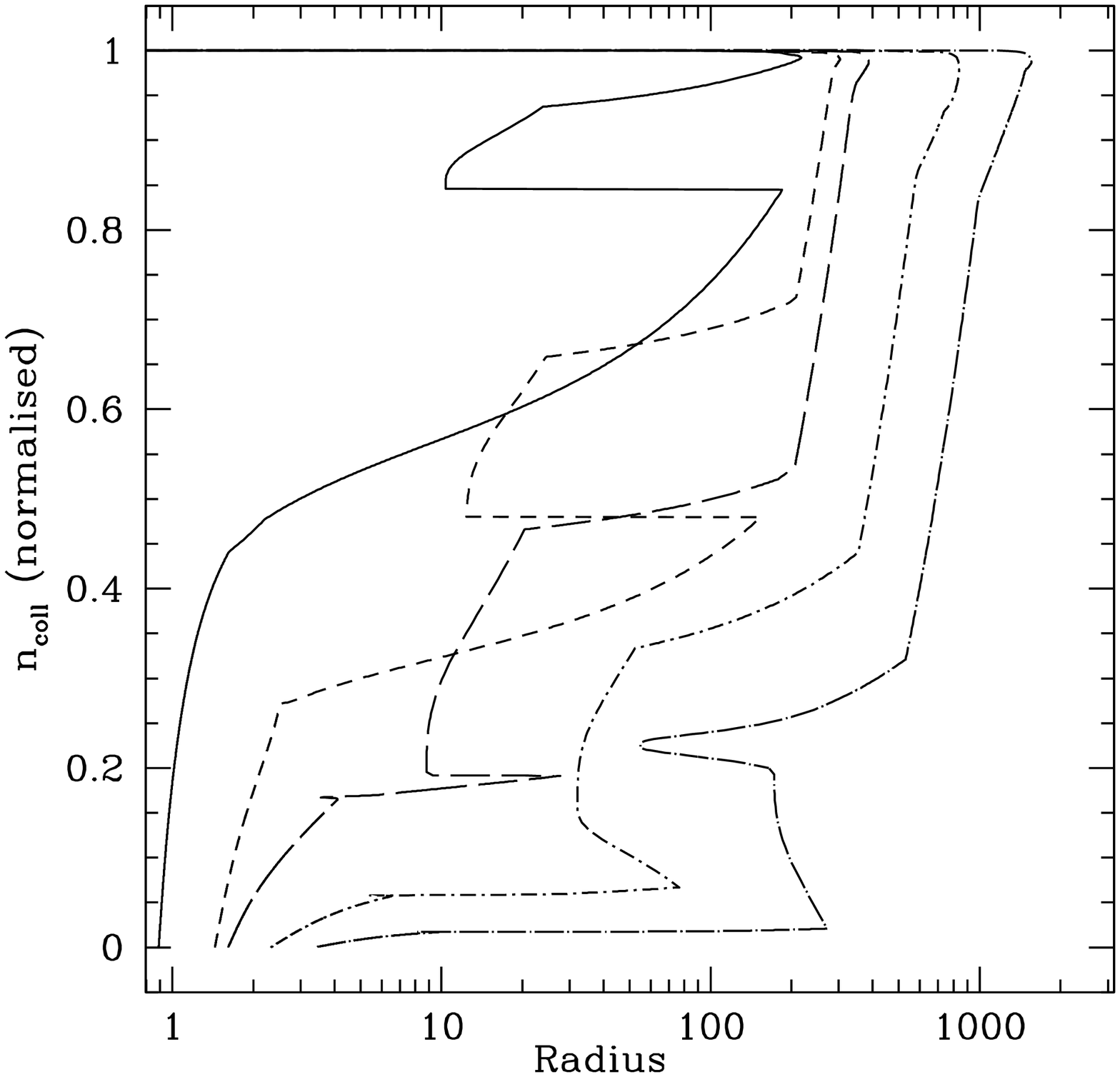}
\caption{The distribution of collisions, $n_{\rm coll}$ (normalised),
during the lifetimes of stars of various masses, plotted as a
function of the stellar radius (solar units), which emphasises different evolutionary
stages. The solid line represents a $1\,$M$_\odot$ star, the short-dashed
curve represents a $1.5\,$M$_\odot$ star, the long-dashed curve corresponds
to a $2\,$M$_\odot$ star, the short-dashed, dotted line represents a $4\,$M$_\odot$
star and the long-dashed, dotted curve corresponds to an $8\,$M$_\odot$ star. 
We have
assumed that all impinging stars are solar mass and solar radius, and
have adopted $\sigma=183.13\,$km/s, appropriate for a galactocentric
radius of $0.17\,$pc.}
\label{fig:zigzag2}
\end{figure}

The calculations shown in Figure~\ref{fig:timesc} have used an
instantaneous radius for the stars. Such calculations lose their
meaning, however, if the stellar radius changes on time-scales shorter
than the collision time-scale. We now
calculate the number of collisions that a star will have by
integrating the instantaneous collision rate over its lifetime:
\begin{equation}
n_{\rm coll}= \int_\tau \Gamma_{1} \, dt.
\label{eqn:zigzag}
\end{equation}

\begin{table*}
\begin{tabular}{lllllll}
\hline\hline\\

Mass [$M_\odot$] & $r$ [$pc$] & Profile &\multicolumn{4}{c}{$n_{\rm coll}$} \\
&&& LIFE &  MS & K$\,<15$ (A$_{\rm K}=3$) & K$\,<15$ (A$_{\rm K}=4$)\\
\hline\hline\\
1   & 0.12 & G96 & 0.741 & 0.292 & 0.278 & 0.229 \\
    & 0.12 & cusp & 1.74 & 0.686 & 0.653 & 0.538  \\
    & 0.17 & G96 & 0.549 & 0.242 & 0.181 & 0.148  \\
    & 0.17 & cusp & 0.894 & 0.394 & 0.295 & 0.242  \\
\hline\\
1.5 & 0.12 & G96 & 0.550 & 0.129 & 0.320 & 0.272 \\
    & 0.12 & cusp & 1.29 & 0.302 & 0.751 & 0.640 \\
    & 0.17 & G96 & 0.387 & 0.106 & 0.208 & 0.176 \\
    & 0.17 & cusp & 0.631 & 0.172 & 0.339 & 0.286 \\
\hline\\
2   & 0.12  &G96 & 0.549 & 0.077 & 0.340 & 0.313 \\
    & 0.12  &cusp & 1.29 & 0.182  & 0.797 & 0.734  \\
    & 0.17  &G96 & 0.379 & 0.0636 & 0.219 & 0.201 \\
    & 0.17  &cusp & 0.619 & 0.104 & 0.358 & 0.327 \\
\hline\\
4   & 0.12 & G96 & 0.533 & 0.0246 & 0.510 & 0.508 \\
    & 0.12 & cusp & 1.25 & 0.0576 & 1.19 & 1.19 \\
    & 0.17 & G96 & 0.354 & 0.0205 & 0.333 & 0.333 \\
    & 0.17 & cusp & 0.576 & 0.0335 & 0.524 & 0.542 \\
\hline\\
8   & 0.12 & G96 &  0.909 & 0.0116 & 0.900 & 0.898 \\
    & 0.12 & cusp & 2.13 & 0.0272 & 2.11  & 2.11  \\
    & 0.17 & G96 & 0.589 & 0.00990 & 0.582 & 0.580 \\
    & 0.17 & cusp & 0.961 & 0.0161 & 0.948 & 0.945 \\
\hline\hline\\ 
\multicolumn{7}{c}{Durations [yr]}\\
\hline\\
1 &&& $1.25\,\times\,10^{10}$& $1.10\,\times\,10^{10}$ & $5.22\,\times\,10^7$&
$1.78\,\times\,10^7$\\

1.5 &&& $3.10\,\times\,10^{9}$ & $2.73\,\times\,10^{9}$ &
  $4.69\,\times\,10^7$ & $1.09\,\times\,10^7$\\

2 &&& $1.50\,\times\,10^9$& $1.17\,\times\,10^9$ & $2.73\,\times\,10^7$&
$4.53\,\times\,10^6$\\

4&&& $2.15\,\times\,10^8$ & $1.79\,\times\,10^8$ & $3.56\,\times\,10^7$ &
$3.52\,\times\,10^7$\\ 

8 &&& $4.25\,\times\,10^7$& $3.72\,\times\,10^7$& $9.61\,\times\,10^6$&
$5.21\,\times\,10^6$\\
\hline\\
\end{tabular}
\caption{Phase durations and number of collisions, $n_{\rm coll}$, during
various evolutionary phases 
for stars of masses 
$1\,$M$_\odot$, $1.5\,$M$_\odot$, $2\,$M$_\odot$, $4\,$M$_\odot$ 
and $8\,$M$_\odot$ at 
galactocentric radii of $0.12\,$pc and $0.17\,$pc. Profile
denotes stellar number-density
profile used: G96 refers the density of Genzel \etal (1996) and cusp 
refers to a density profile of $\rho \propto r^{-1.8}$,
set equal to its counterpart at $r\ge 0.38\,$pc. We have assumed that the mean
mass of the GC stars, {\rm ${\bar M_2}$}, is 
$1\,$M$_\odot$ but note that, approximately, $n_{\rm
coll}\propto {\rm 1/{\bar M_2}}$.  LIFE represents the 
life-span of the star from ZAMS to AGB termination, 
MS represents the main-sequence duration
and K$\,<15$ represents the time for which the stars are brighter
than K=15 for a given extinction. Evolutionary models of stars are
from Z=0.02 models from Pols \etal (1999).}
\label{tbl:time}
\end{table*} 
In this way we obtain Table~\ref{tbl:time} in which we have restricted
$\tau$ to interesting parts of the star's evolution as well as the
entire lifetime, for stars of various masses. 
 We see that $n_{\rm coll}$, when evaluated over the
lifetime of a star, can be a substantial fraction of unity. 
For sufficiently low radii, and
preferably with a steep cusp density profile, $n_{\rm coll}$ is about
one. The more-massive stars are more luminous than the $1\,$M$_\odot$ star
and so spend a larger fraction of their lives as K$\,<15$ magnitude
stars. During this period they reach larger radii than $1\,$M$_\odot$
stars (Figure~\ref{fig:HR1}) and so we expect them to undergo greater
proportions of collisions as detectable giants. There are a number of
uncertainties in $n_{\rm coll}$. Uncertainties in the stellar
number-density profile are discussed in $\S$2.4. The value of the mean
stellar mass (${\bar M_2}$) is uncertain. In Table~\ref{tbl:time} 
we set the number of impacting stars to be $n_2(r)=\rho_0n(r)/{\bar
M_2}$, where $n(r)$ is given by Equation~(\ref{eqn:ndens}), and where
${\bar M_2}=1\,$M$_\odot.$ Thus, via Equation~(\ref{eqn:gam}),
$\Gamma_1\propto 1/{\bar M_2}$ (ignoring gravitational focusing). 
A decrease in ${\bar M_2}$ to, say, $0.\,7M_\odot$, increases 
$\Gamma_1$ by a factor $\sim 1.4$. 
$n_{\rm coll}$ is also sensitive to the radius-time
behaviour of the stars. Changing the
mass-loss rate by a factor of 10 in the models of Pols \etal (1999)
produces a factor of two change in $n_{\rm coll}$. 

The relative likelihood of collisions during different phases of the
stars' evolution are shown in
Figure~\ref{fig:zigzag2}. We see that one of the stars in
single-single stellar collisions in the GC will most likely be a giant
star (provided that it is sufficiently old to have turned off the main
sequence). In terms of collision
frequency, the AGB becomes 
increasingly predominant over the first giant branch 
for stars of higher mass.
Only a minority fraction of target
stars are on the main sequence when hit by an impactor. 
Although the main-sequence
phase is much longer lived than the giant phase, more two-body 
collisions in the GC will
occur whilst the target star is a giant because of its much greater
radius compared to the main-sequence radius.
In fact the number of 
target stars on the main sequence is $\lo 25\%$, with the exception of
$45\%$ for a $1\,$M$_\odot$ star. The importance of the main-sequence
phase decreases
with increasing stellar mass. The horizontal branch also generally
dominates over the main sequence in competition 
for collisions: it is shorter-lived but the star is considerably
larger on the horizontal branch than when on the main sequence.

\section{Criteria for destruction of giants}

To determine whether stellar collisions may destroy giant stars, one
needs to know how much envelope mass a giant must lose before it is
destroyed. Strictly
speaking, to satisfy the observed paucity we must strongly reduce the
K-magnitude of the brightest giants. Considering mild encounters where
the mass lost is insufficient to alter the giant's long-term evolution, we
actually expect a transient {\it increase} in the stellar
K-luminosity: the impactor loses some of its kinetic energy to the
envelope which consequently expands somewhat, becomes redder and
therefore brighter in the K-band. The envelope will then settle back
to normal on its thermal time-scale ($\sim\,$few-hundred years). To
explain the paucity of giants, collisions must therefore destroy the
giant, \ie they must cause large amounts of envelope loss. 
The actual amount of mass loss required as a direct result of a collision 
is uncertain but is likely to be most if not nearly all of 
the envelope (Eggleton, Tout, private communications). A second way to
destroy the giant is via indirect mass loss after a collision produces
a common envelope system. In
such systems, the binding energy released by the impactor and core
spiralling in towards each other may unbind the envelope. To form a
common envelope system in a hyperbolic collision requires the impactor
to deposit at least $\mu$$V_\infty^2$/2 of its orbital energy into the
giant but need not necessarily produce copious mass loss.  
Thus to
destroy a giant, a collision must involve either immediate, large mass
losses or the formation of a  common envelope system which then evolves to
expel the envelope ($\S$7).

\section{Approach to hydrodynamical simulations}

We simulated encounters involving giant stars using a
smoothed-particle hydrodynamics (SPH) code. For a discussion of SPH, see
Benz (1990). The impactors encountering the star were point masses which were
adequate to simulate either main-sequence stars or compact remnants
such as white dwarfs, neutron stars or stellar black holes because of
the size contrast between these objects and the
giant's envelope. 

\subsection{Collisions performed}

We performed collisions on two red giants. The radii and luminosity of
the stars were selected so that their positions on a HR diagram
coincided with those of the missing bright giants (see
Figure~\ref{fig:HR1}). 
One giant had mass $2\,$M$_\odot$, radius $94\,$R$_\odot$ and core mass
$0.52\,$M$_\odot$. The other giant had
mass $8\,$M$_\odot$, radius
 $243\,$R$_\odot$ and core mass $2.5\,$M$_\odot$. In each case we
represented the core by a point mass. The masses of the
two stars were selected so that we sampled both ends of the
likely mass range of the missing giants (see $\S$2).  Both giants
were based upon
models of temperature and density profiles discussed in Pols \etal (1995).

Most simulations were performed with $\sim\,8500$ particles. A number of
simulations were run with $16\,\times\,10^3$ and $36\,\times\,10^3$ 
particles in order to quantify resolution effects (see
$\S$6.3). A wide range of impactor
masses were selected. For the $2\,$M$_\odot$ giant we used $0.6\,$M$_\odot$, $1\,$M$_\odot$,
$1.4\,$M$_\odot$, $2\,$M$_\odot$, $4\,$M$_\odot$ and $8\,$M$_\odot$ impactor
masses. Our results
suggested that for the $8\,$M$_\odot$ giant, results from impacts with a
$1\,$M$_\odot$ impactor would be quantitatively similar to  the outcomes 
of the giant's encounters
with $0.6\,$M$_\odot$ and $1.4\,$M$_\odot$ impactors, 
so for the more-massive giant
we simulated collisions involving only $1\,$M$_\odot$, $2\,$M$_\odot$, $4\,$M$_\odot$ and
$8\,$M$_\odot$ impactors.

For each giant-impactor pair, we ran $\sim\,16\,-\,25$ simulations
distributed on a $\vinf\,-\,\rmin$ grid. A number of simulations
were also performed very near the $\vinf$ and $\rmin$ axes in order to
quantify our extrapolations from the grid to low $\vinf$ and low
$\rmin$ values (see $\S$6.3). In total approximately 200 runs were
performed.

\subsection{Quantifying collision outcomes}

To determine whether a collision destroyed the giant, we measured the
mass loss, $\Delta M$, searching for mass losses of most 
 of the envelope ($\S$4). To determine $\Delta M$, the enthalpy
 of each particle was
evaluated relative to both the impactor and the giant, using a
similar method as prescribed in Rasio \& Shapiro (1991). For each
collision we ascertained the amount of mass remaining bound to the
giant, the amount that had become bound to the impactor and the
amount of escaping matter. We found that
$\Delta M$ settled down within a sound-crossing time or two after
 a collision. 

A number of collisions involved dissipation of enough of the
impactor's orbital energy for the impactor to become bound to the giant. Such
systems are important as they potentially lead to the destruction of
the giant via a common envelope phase ($\S$6.2). To delimit the region
of the $\vinf\,-\,\rmin$ plane in which two impactors became bound, we
measure the fractional change in their orbital energy, $\deltaee$. We
define $\Delta E$ such that $\deltaee=1$ when the impactor just
becomes bound to the giant. The
initial orbital energy is simply $\mu\vinf^2/2$; the final orbital
energy is calculated from the
masses, positions and velocities of the two centres of mass. 

For some of the bound systems formed, the impactor remained inside the
envelope, \ie a common envelope system had 
effectively been formed immediately. 
In such systems we do not have two distinct objects to
which we can assign fluid particles. Our algorithm is thus unsuitable
for such outcomes and so $\Delta M$ and $\deltaee$ were not measured
for these outcomes.

\section{Results}

\begin{figure}
\psboxto(\hsize;0cm){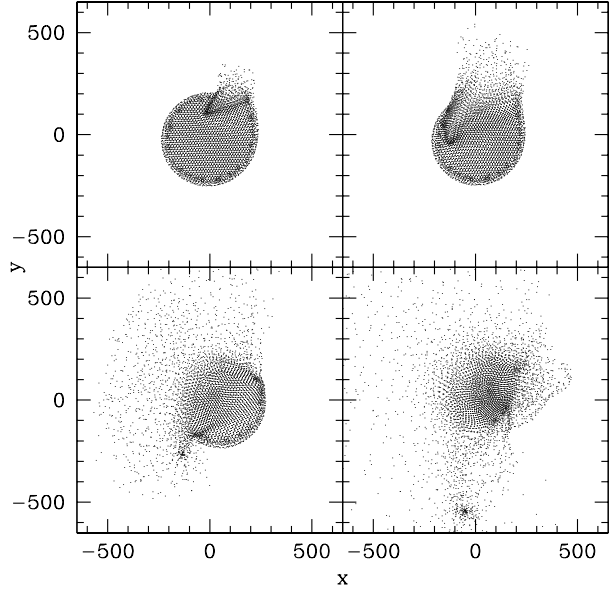}
\caption{Time sequence of fluid-particle plots of an SPH 
simulation between an $8\,$M$_\odot$ giant and a $1\,$M$_\odot$ impactor
(centre-of-mass frame), with $\vinf=50\,$km/s and $\rmin=100\,$R$_\odot$.
Here we take a slice in the plane of the
collision. The trajectory of the impactor is clearly deflected; this
is due to the giant's mass enclosed in its orbit. The top-left and
top-right 
portions in particular show the shock caused by the impact. 
Units of x and y are solar radii. }
\label{fig:partplot}
\end{figure}

The impactor hits the giant supersonically, shocking the envelope, as
can be seen in  Figure~\ref{fig:partplot}.  Due to the giant's
gravity, the impactor is pulled around the core somewhat; the amount
of deflection depending on the $\vinf$ of the encounter. 
Almost all of the escaping particles had speeds in
excess of the local escape speed ($V_{\rm esc}=\sqrt{
2GM_1/R}$ for $R\geq R_1$, where $R_1$ is the giant's radius), 
demonstrating that the escaping
fluid was energetically unbound by virtue of its kinetic energy rather
than its thermal energy. Some
fluid had speeds up to four times that required to escape and thus
carried off excess energy thereby reducing the mass loss. Given that
mass losses are generally low ($\S$6.1), this implies
that deep collisions are required to remove envelope material located
close to the core. 

Both mass loss from the giant and
orbital energy loss from the two stars increase with
increasing impactor mass and with decreasing $V_\infty$, indicating 
the dominance of
time-dependent gravity over shocks as the energy-transfer
mechanism. Mass loss and orbital energy loss also increase with
decreasing $\rmin$.

We divide the collision outcomes into two categories: bound and
unbound giant-impactor pairs (after collision). A bound system is
defined by $\Delta E/E\geq 1$ (see $\S$5.2). Both mass loss
(Figure~\ref{fig:mass2}) and
orbital energy loss seemed to vary smoothly across the boundary
(Figure~\ref{fig:de1}).

\subsection{Mass loss}

We reiterate that the instantaneous mass loss increases with impactor
mass and falls with increasing $\vinf$ and $\rmin$. For two given
colliding species, this
implies that we would observe maximal mass loss for collisions of
lowest $\vinf$ {\it and} lowest $\rmin$. Those particular encounters, 
however, tended to directly produce common envelope systems (see
$\S$6.2) for which we do not measure the mass
loss due to the collision. 
Consequently the subsequent comments concerning $\Delta M$ are
limited to outcomes that are not common envelope systems and that
constitute the larger portion of our  data.

\begin{figure}
\psboxto(\hsize;0cm){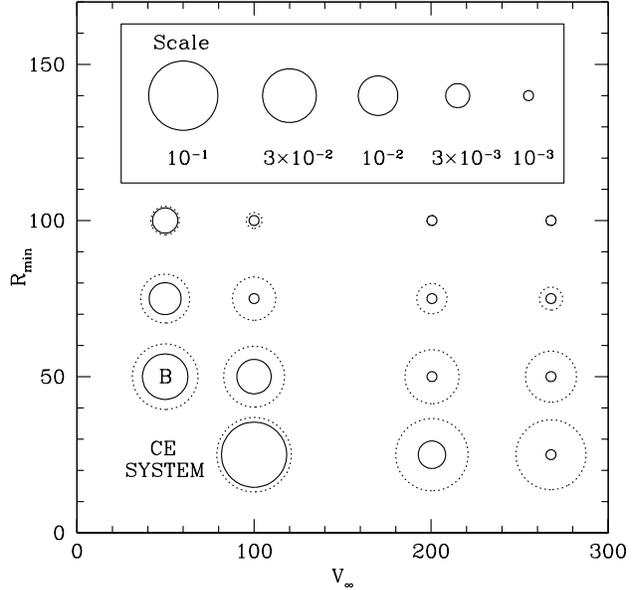}
\caption{Impactor-accreted mass (unbroken circles) and mass lost from the
system in collisions (broken circles) between the $2\,$M$_\odot$ giant and a
$1\,$M$_\odot$ impactor, as a function of $V_\infty$ [km/s] and $R_{\rm min}$
[R$_\odot$]. The symbol ``B'' represents that the collision product is
a binary, and ``CE SYSTEM'' denotes the direct formation of a common
envelope system. The size of the circles is proportional to the log
of the mass. Masses are in solar units.}
\label{fig:mass2}
\end{figure}

Collisions of both giants with impactors of mass $\sim\,1\,$M$_\odot$ 
involved small mass losses, $\lo 0.1\,$M$_\odot$ (and usually much
less, see Figure~\ref{fig:mass2}). 
In fact, for the $8\,$M$_\odot$ giant and for $\vinf\go 100\,$km/s, the
mass losses were so small that too few SPH particles escaped to take a
reliable measurement of $\Delta M$. Even deep impacts ($\rmin \sim\,
R_1/4$) were not devastating for either giant, although the mass
lost from the $2\,$M$_\odot$ giant rose to $\sim\,0.1\,$M$_\odot$. 

If the bulk of the population in the GC are stars of masses 
$\lo$1$M_\odot$, then
the paucity of bright giants cannot be explained by instantaneous mass
loss due to two-body collisions because the vast majority of
collisions with $1\,$M$_\odot$ impactors will involve mass losses much
less
than the envelope mass. 

Mass losses were higher for more-massive impactors because of the
greater gravitational forces involved. We discuss here
the effects of $8\,$M$_\odot$ impactors. Encounters involving our $8\,$M$_\odot$ giant
showed mass losses of $\sim\,0.1\,-1\,$M$_\odot$ for
$\rmin \go 50\,$R$_\odot$. Collisions involving the $2\,$M$_\odot$ giant with
$\rmin>25\,$R$_\odot$ gave mass losses $\lo 0.5\,$M$_\odot$,
dropping to a few $\times 10^{-2}\,$M$_\odot$ 
for fast ($250\,-\,450\,$km/s) grazing
encounters. Deeper impacts had much stronger effects for both
giants. For our $8\,$M$_\odot$ giant, collisions with $\rmin<50\,$R$_\odot$ gave rise
to mass losses of about $3\,$M$_\odot$ for $\vinf\go 150\,$km/s (our slower
collisions gave rise to common envelope systems). For the
$2\,$M$_\odot$ giant, encounters of
$\rmin\lo 25\,$R$_\odot$ involved mass losses approaching the entire
envelope mass. It is possible that low $\rmin$ ($\lo 30\,$R$_\odot$)
collisions with
massive impactors could destroy the giants. Such collisions, however,
will be rare; considering the low $\rmin$ alone places {\it upper
limits} of frequencies of $5\%$ of all giant-$8\,$M$_\odot$ impactor collisions
for the $8\,$M$_\odot$ giant and $20\%$ for the
$2\,$M$_\odot$ giant, assuming that all collisions occur at
$\vinf=\sqrt{2}\sigma(r)$ and that $r=0.2\,$pc. 
We note that in $\S$3 we showed that a star
could expect at most about one collision before becoming a remnant; 
most collisions will occur at
too high $\rmin$ to destroy the giant. Moreover, to
destroy all giants in this manner would further require the GC to
contain a large proportion of impactors of
masses $\go 8\,$M$_\odot$. 
Such a population seems quite unlikely, unless, perhaps,
the GC core is dominated by massive black holes which have diffused
inwards from higher galactocentric radii (see $\S$2 and Morris 1993).   

In summary, mass losses due to most two-body collisions 
are likely to be small
($\lo 0.1\,$M$_\odot$), and though very massive impactors might be able to
destroy the giants, the deep-envelope collisions required will make
such encounters rare, unless the GC contains an exotic population of
numerous massive stars which is itself unlikely.

\subsection{Formation of bound systems}

For collisions of sufficiently low $\vinf$ and $\rmin$, the impactor lost
enough kinetic energy to become bound to the giant. Such collisions
will almost always be physical collisions and we expect the
resulting orbits' periastrons to remain inside the giant. If the
apastron of the orbits were also within the giant's radius then a
common envelope system had effectively immediately been
formed. Systems with apastrons outside the giant's envelope were very
eccentric and had
periods of a few years to centuries. Thus the time-scale for further
collisions is much shorter than the giant's evolution time-scale. We
expect each subsequent encounter to dissipate 
more of the orbital energy until the
apastron falls within the giant's radius (or the envelope puffs up
enough to absorb the impactor's orbit), producing a common envelope
system. 

Common envelope systems are of great interest here as their evolution
may quickly destroy the giant. The impactor and giant's core spiral
towards each other, transferring angular momentum and energy to the
envelope. If a sufficient quantity of the
potential energy between the
impactor and giant's core is deposited into the envelope, the entire envelope
may be expelled, leaving a tight binary. Alternatively, the impactor
may merge with the core before envelope ejection is complete,
terminating the common envelope phase. The outcomes of common envelope
evolution involving either of the two giants considered here are 
discussed in $\S$7, here we
calculate the fraction of collisions that produce
them. 

\begin{figure}
\psboxto(\hsize;0cm){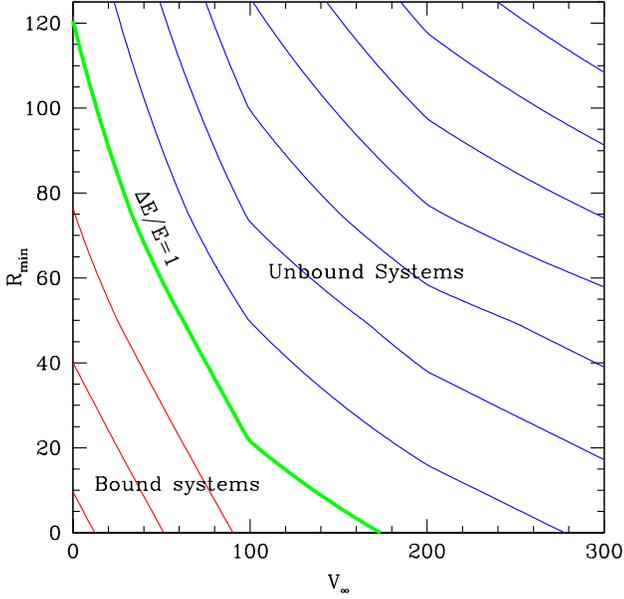}
\caption{$\Delta E/E$ contour plot for collisions involving the
$2\,$M$_\odot$ giant and a
$1\,$M$_\odot$ impactor, as a function of $V_\infty$ [km/s] and $R_{\rm min}$
[R$_\odot$]. The contours are logarithmically spaced by 0.5, the bold
line represents $\Delta E/E=1$ \ie the threshold of
bound-system formation. Contours to the bottom-left of this line have
$\Delta E/E>1$ so the bottom-left region represents the region 
where bound systems are formed. Collisions of the upper-right
portion leave the two stars unbound.}
\label{fig:de1}
\end{figure}

\begin{figure}
\psboxto(\hsize;0cm){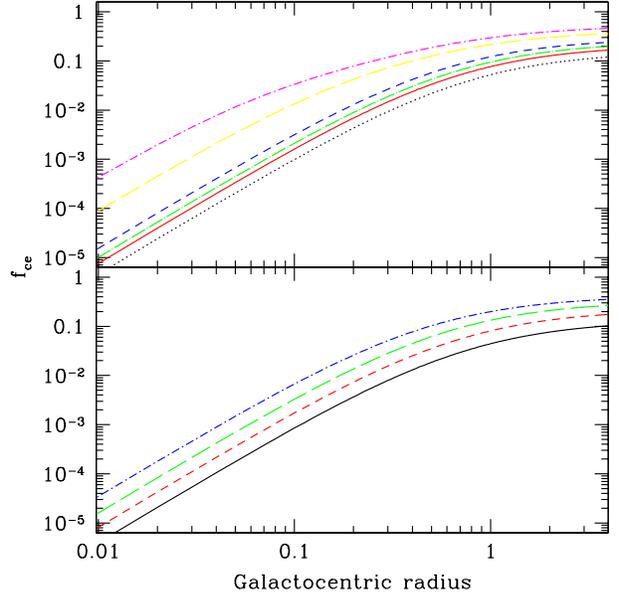}
\caption{The fraction of encounters that produce bound systems as a
function of galactocentric radius (in parsecs): the
$8\,$M$_\odot$ giant (lower half) and the $2\,$M$_\odot$ giant (upper
half). Different curves reflect different impactor masses with curves
of increasing height representing increasing impactor mass. For the
$8\,$M$_\odot$ giant, the impactor masses are $1\,$M$_\odot$, $2\,$M$_\odot$,
$4\,$M$_\odot$ and
$8\,$M$_\odot$. The same notation has been used for the $2\,$M$_\odot$ giant, except
we also show curves for $0.6\,$M$_\odot$ impactors (dotted line) and $1.4\,$M$_\odot$
impactors (dot-dashed line). We have used Equation~(\protect\ref{eqn:sig1}) to
describe $\sigma(r)$.}
\label{fig:frac}
\end{figure}

Figure~\ref{fig:de1} shows the dependence of the fractional change
in pre-collision orbital energy on $\vinf$ and $\rmin$ for a certain
giant-impactor pair. Bound systems are produced in the bottom-left of
the Figure and are delimited from unbound systems by the
$\Delta E/E=1$ contour (which we now denote as $R_{\rm b}(\vinf)$). 
As impactor mass increases, the delimiting contour moves toward the
top-right portion of the Figure. Values of $\Delta E/E$ between data
points are obtained by linear interpolation of a fitted 2D exponential function
$\Delta E/E(\vinf,\rmin$). Note we do not measure $\Delta E/E$ for common
envelope systems, rather, we extrapolate from the other data.
The fraction of collisions, $f_{\rm ce}$, that fall
underneath the $R_{\rm b}(\vinf)$ curve depends on the 
likelihood of having a given velocity ($\vinf$) and
therefore on $P(\vinf)$, the velocity distribution function, which we
take to be Maxwellian ($\S3$). For a given $\vinf$, $f_{\rm ce}$ 
depends on the proportion of collisions that have
$\rmin \leq R_{\rm b}(\vinf)$ and thus depends on the amount of gravitational
focusing (Equation~[\ref{eqn:gsig}]):
\begin{equation}
f_{\rm ce} = { \int_{0}^{\infty} \Sigma_{\rm g} \left( V_{\infty},
R_{\rm min}=R_{\rm b} (V_{\infty})\right) V_{\infty}
P(V_{\infty}) \, dV_{\infty} \over \int_{0}^{\infty}
\Sigma_{\rm g}(V_{\infty}, R_{\rm min}= R_1) V_{\infty} P(V_{\infty}) \,
dV_{\infty}}.
\label{eqn:frac}
\end{equation}
The dependence of
$\sigma(r)$ on position within the GC exerts a galactocentric radial
dependence on $f_{\rm ce}$ and is shown in Figure~\ref{fig:frac}. 

We see that $f_{\rm ce}$
increases with impactor mass, which follows because the increased
gravitational interaction allows bound systems to be formed at both 
higher speeds and at higher $\rmin$. Figure~\ref{fig:de1} shows that
for increasing $\vinf$, the largest $\rmin$ that produces a bound system
decreases. As galactocentric radius decreases,
the median $\vinf$ increases and so we expect $f_{\rm ce}$ to decrease 
rapidly inside the core radius ($\sim\,0.38\,$pc) and to flatten at
high radii in
accordance with the nature of the velocity dispersion
(Equation~[\ref{eqn:sig1}]), as illustrated by Figure~\ref{fig:frac}. 
Within the core radius the fraction of bound systems formed is only a
few percent
($\sim\,1\,-10\%$ at the core radius). At a few parsecs,
$f_{\rm ce}\sim\,10\,-\,40\%$ for the range of impactor masses
considered. The
large $f_{\rm ce}$ at
high galactocentric radii is due to the low local velocity dispersion.
 
We note that $f_{\rm ce}$ is substantially less than one, even for
collisions between the $8\,$M$_\odot$ giant with $8\,$M$_\odot$
impactors at high galactocentric radii (\ie low speeds). Since $f_{\rm
ce}$ is the fraction of physical collisions that lead to the capture
of the impactor, for tidal capture to occur we require $f_{\rm
ce}>1$. Thus tidal capture by giants similar to those
discussed here will be extremely rare in the galactic centre.

\subsection{Resolution and error considerations}

Three sources of error in our data were considered: the accuracy of
extrapolation of $\deltaee$ to regions near the $\vinf$ and $\rmin$
axes (Figure~\ref{fig:de1}), errors in energy conservation in
individual SPH runs and errors due to resolution effects (\ie number of
particles used) in our simulations. 

To assess errors in $\deltaee$ extrapolation we performed a number of
simulations near the $\vinf$ and $\rmin$ axes and compared the local
extrapolated 
$\deltaee$ value to that obtained from the extra runs. We then locally
refitted $R_1(\vinf)$ and re-calculated $f_{\rm ce}$ to evaluate its
error. Errors in $f_{\rm ce}$ due to energy conservation in individual
runs were examined by estimating their effects on the local $\deltaee$
value and hence on the locus of the $R_{\rm b}(\vinf)$ contour, producing an
error in $f_{\rm ce}$. In both cases the error in $f_{\rm ce}$ 
for all giant-impactor pairs throughout most of the GC is within
$\pm 30\%$. Slightly over-widely spaced grids for the $2\,$M$_\odot$ giant
and  $2\,$M$_\odot$, $4\,$M$_\odot$ and $8\,$M$_\odot$ impactors led to larger
errors at low galactocentric radii, but the net effect is likely to be
a mild overestimation (within a factor of two) of $f_{\rm ce}$ for any
galactocentric radius outside of the central few hundredths of a
parsec. These errors do not weaken our conclusions; indeed,
overestimation of $f_{\rm ce}$ actually acts to strengthen our
conclusions (see $\S$8).

We also estimated errors in $\Delta M$ and $\deltaee$ (and hence
$f_{\rm ce}$) due to resolution, \ie using a limited number of particles in our
simulations.  We repeated a quantity of collisions involving the
$8\,$M$_\odot$ giant with $16\,\times 10^3$ and $36\,\times 10^3$ particles.
The quantity of interest (\eg $\Delta M$) was evaluated as a function
of the particle separation. 
Extrapolation to zero spacing (\ie infinite resolution)
obtains the best estimate of the quantity. 
We found that our mass losses became generally good estimates for 
$\Delta M \go\,$ few$\,\times\,0.01\,-\,0.1\,$M$_\odot$, depending on impactor
mass. The resolution
errors in $\deltaee$ caused uncertainties in $f_{\rm ce}$ that were
generally small and never large enough to affect our conclusions. 
Since the $2\,$M$_\odot$ giant shows, to some extent, similar scaling to
the $8\,$M$_\odot$ giant (\eg number of particles per fractional radius),
we expect similar resolution errors.  

The uncertainties considered here do not affect our conclusions, and so we
have not included them in our calculations.

\section{Evolution of binary systems formed}

We now consider the outcomes of the evolution of common envelope
systems formed by our two-body encounters. Within the envelope, the
core and impactor orbit each other in ever tightening orbits, their
potential energy being converted into their orbital kinetic energy and
thence deposited in the envelope. If the envelope gains enough energy,
it will become unbound. We take the binding energy of the envelope to
be (de Kool 1990):
\begin{equation}
E_{\rm env} = \frac{G(M_{\rm c} + M_{\rm env})M_{\rm env}}
{\lambda R_{\rm env}}
\label{eqn:eenv}
\end{equation}
where $M_{\rm c}$ is the giant's core mass, $M_{\rm env}$ and
$R_{\rm env}$ are respectively the envelope's mass and radius and we
take $\lambda$ to be 0.5. The change in binding energy between the
impactor and the giant from the onset of common envelope
evolution to its termination when the entire envelope has been expelled is given by:
\begin{equation}
\Delta E_{\rm fi} = \frac{GM_{2}M_{\rm c}}{2d_{\rm f}} -
\frac{GM_{2}(M_{\rm c}+M_{\rm env})}{2d_{\rm i}}
\label{eqn:efi}
\end{equation}
where $M_{2}$ is the impactor mass, $d_{\rm i}$ and $d_{\rm f}$ are the
initial and final semi-major axes.
By assuming all the envelope is removed with
some efficiency, $\alpha$, then
\begin{equation}
E_{\rm env}=\alpha \Delta E_{\rm fi}.
\label{eqn:ealpha}
\end{equation}
Thus we may calculate the final separation between the impactor and
core:
\begin{equation}
d_{\rm f} = M_{\rm c} \left(\frac{2E_{\rm env}}{G \alpha M_{2}}
+ \frac{M_{\rm c} + M_{\rm env}}{d_{\rm i}} \right)^{-1}.
\label{eqn:df}
\end{equation}
The following results are insensitive to $d_{\rm i}$ since $d_{\rm i}$
must be very deep in the envelope before the second right-hand term
becomes comparable to the first.
Numerical modelling of common envelope evolution suggests
that $\alpha \sim\,0.24\,-\,0.6$ (Taam \& Bodenheimer 1989, Sandquist \etal
1998); we take $\alpha=0.4$.
 Using the parameters for our giants (see $\S$5), for both
giants we find that $d_{\rm
f}$ takes values between $\sim\,0.8-4\,$R$_\odot$ 
for impactor masses in the range $0.6\,$M$_\odot$ to $3\,$M$_\odot$. 
$d_{\rm f}$ is similar for both giants because the greater energy required to
unbind the $8\,$M$_\odot$ giant's envelope is offset by the greater energy
released by an impactor falling towards its more massive core. 

\begin{figure}
\psboxto(\hsize;0cm){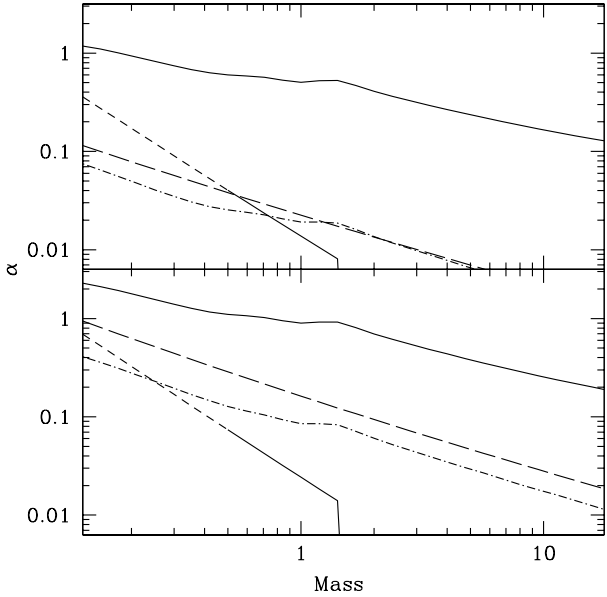}
\caption{Envelope-ejection efficiency, as a function of impactor mass
[$M_\odot$], for which the envelopes
of our giants may be expelled before the impactor sinks deep enough
into the giant that the impactor or the core fills its Roche lobe or
the density of the giant becomes comparable to that of the impactor. 
The lower box represents the $8\,$M$_\odot$ giant and the upper box
represents the $2\,$M$_\odot$ giant. For each giant, the upper-most
(solid) line corresponds to Roche-lobe filling by ZAMS impactors. The
lower, solid line represents Roche-lobe filling by stellar remnants; 
the broken region  represents remnants that, if they exist, 
must be produced in a
binary since their solitary progenitors live longer than a Hubble
time. The almost vertical drop
in the remnant curve corresponds to the Chandrasekhar mass limit
($\sim\,1.44\,$M$_\odot$), above which the remnants are neutron stars (or
black holes for even larger remnant masses) and, effectively, are
point masses requiring very low $\alpha$ to eject the envelope. The
long-dashed line represents Roche-lobe filling of the giant's core and
the dot-dashed line represents the efficiency at which the ZAMS star
falls deep enough into the envelope that the local density matches its own.}
\label{fig:rlvm}
\end{figure}

We have thus far assumed that the core and impactor are point
masses. Their sizes, however, can be 
important and can lead to termination of the common envelope phase
before total ejection of the envelope. Either the impactor or the core
may fill their Roche lobes or, alternatively, the impactor may fall so
deep into the giant
that the local density is comparable to its own. In the following
calculations we find the former almost always occurs before the latter. 
Either way, the impactor is no longer distinct from its surroundings
and no further potential energy may be liberated. The occurrence of such
``premature'' termination depends on the size of the impactor 
and the other parameters in
Equations~\ref{eqn:eenv}-\ref{eqn:df}. We ascertain whether envelope
ejection will occur in the following way. Two species of impactor are
considered: ZAMS stars and stellar remnants. The radii of the former
are taken from Tout \etal (1996). A simple mass-radius relation for
white dwarfs suffices for this discussion: $R\propto M_1^{-1/3}$
($M_1<1.4\,$M$_\odot$), scaled to $0.0248\,$R$_\odot$ at $1\,$M$_\odot$.
More-massive remnants (neutron stars) are given a size of 10km. We
calculate the impactor's Roche-lobe size using Eggleton (1983):
\begin{equation}
{R_{{\rm L1}} \over d_{\rm f}}={0.49q^{2 \over 3} \over
0.6q^{2 \over 3}+ln(1+q^{1 \over 3})}
\label{eqn:roche}
\end{equation}
where $q=M_2/M_{\rm c}$. Finally, we used the density profiles of our
giant models (see $\S$5) to determine the giant's radius at which the
density was comparable to that of the impactor's density (derived from
its mass and the appropriate mass-radius relation). Our results are
shown in Figure~\ref{fig:rlvm}.

We see that for $\alpha \sim\,0.5$, ZAMS stars of masses $\go 1\,$M$_\odot$
may expel the envelope of the 2$M_\odot$ giant, whereas ZAMS stellar masses
in excess of $2\,$M$_\odot$ are required to expel the $8\,$M$_\odot$ giant's
envelope. Increasing $\alpha$ decreases the minimum mass required.
 We also see that a ZAMS
star will always fill its Roche lobe before penetrating sufficiently
deep into the envelope for the local density to be comparable to its
own. Stellar
remnants should almost always be able to eject the envelopes of either
star, the exception being for low-mass ($\lo 0.4\,$M$_\odot$) remnants with
the $8\,$M$_\odot$ giant. If $\alpha$ is very low, the core will generally 
fill its Roche lobe before a remnant counterpart does, except in the
case of the $2\,$M$_\odot$ giant with remnants of masses $\lo 0.5\,$M$_\odot$. This
is because the $2\,$M$_\odot$ giant's core (of mass $0.52\,$M$_\odot$) is essentially a
white dwarf and 
white dwarfs of lower masses are larger: they will fill their Roche
lobes first. A similar effect is not seen for the $8\,$M$_\odot$ giant: its
core is much larger ($0.25\,$R$_\odot$) than a stellar remnant and tends
to fill its Roche lobe first.

It seems likely that most {\it low-mass} main-sequence stars will not expel
either of the giants' envelopes, particularly that of the $8\,$M$_\odot$
giant. The numbers of common envelope systems formed by two-body
encounters that lead to total envelope expulsion will therefore
be dependent on the remnant fraction in the GC population (see
$\S$2). To be sure that most common envelope systems will expel the
envelope requires a population with a large abundance of stellar
remnants or massive main-sequence stars with few low-mass
main-sequence stars; such a population may exist in the GC if the
star-formation low-mass cut-off is high (\eg $>0.8\,$M$_\odot$) or if
the GC contains an
extra population of neutron stars and black holes (Morris 1993). 

The violence of this method of forming common envelope systems may
partly assist in lowering the impactor-mass limit for full envelope
expulsion. Although the energy deposited in the envelope to make the impactor
bound ($\mu\vinf^2/2$) will dissipate on the envelope's thermal
time-scale of a few-hundred years (perhaps somewhat shorter than
common envelope evolution), the mass loss, however, is
permanent. Both effects tend to lower the envelope's binding
energy. In both cases, though, the effects are strongest for massive
impactors (which are likely to eject the envelope anyway) and are
weakest for low-mass impactors.

\section{Discussion: time-scales for forming common envelope systems}

\begin{figure}
\psboxto(\hsize;0cm){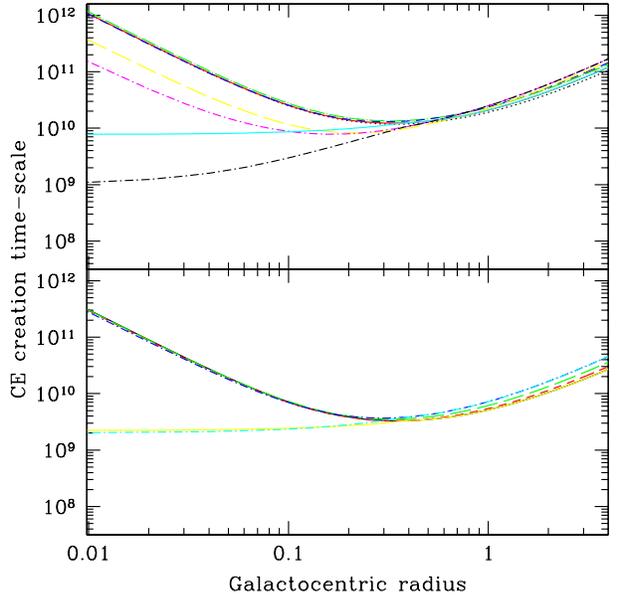}
\caption{Time-scales [years] for the creation of bound 
systems in the galactic centre. The lower graph represents time-scales
for the $8\,$M$_\odot$ giant and the upper graph corresponds to the
$2\,$M$_\odot$ giant. In each case, the solid line represents $1\,$M$_\odot$
impactors, the short-dashed line shows time-scales for $2\,$M$_\odot$ impactors
(this curve is partly concealed by the solid line for the upper
figure), the long-dashed line denotes the $4\,$M$_\odot$ impactors and the
dot-dashed line corresponds to $8\,$M$_\odot$ impactors. In addition, in the
$2\,$M$_\odot$ giant portion the 
dotted curve represents collisions with $0.6\,$M$_\odot$ impactors and the
long-dashed dotted curve shows the bound-system
creation time-scales with $1.4\,$M$_\odot$
impactors. We have used
the velocity dispersion and density profile of  
Genzel \etal (1996). For both target giants, the two less-bold sets
of curves (which point to low time-scales at low
galactocentric radii) correspond to impactor masses of $1\,$M$_\odot$
(solid line) and $8\,$M$_\odot$ (dot-dashed line), 
but we have replaced the density of Genzel
\etal (1996) by the $\rho \propto r^{-1.8}$ profile (see $\S2.4$).}
\label{fig:detime}
\end{figure}

The time-scale for a giant to undergo a two-body collision that forms a bound
system is given by 
\begin{equation}
\tau_{\rm ce}(r)={ 1 \over \Gamma_1({\rm r}) f_{\rm ce}(r)}
\label{eqn:tce}
\end{equation}
where $f_{\rm ce}(r)$ is the fraction of collisions that form bound
systems and  $\Gamma_1(r)$ is the collision rate per giant, given by
Equation~(\ref{eqn:gam}). If all such bound giant-impactor
pairs develop into
common envelope systems (see $\S$6) which then evolve to expel the
envelope, $\tau_{\rm ce}$ is then a lower limit for the destruction of
our giants in the GC, excluding the rare collisions in which the
instantaneous mass loss approached the total envelope mass ($\S$6). 
We minimise $\tau_{\rm ce}$ by assuming that, for a given
impactor mass, $M_2$, {\it all}  stars in the GC (other than our giant)
are of mass $M_2$. This produces Figure~\ref{fig:detime} where we see
that for both giants $\tau_{\rm ce}>10^9$ years,
even if $n\propto r^{-1.8}$.
This time-scale is longer than the
lifetimes of stars of masses $\go 2\,$M$_\odot$.  
Given the rarity of high mass-loss encounters, we
conclude that the paucity of luminous red-giant stars is unlikely
to be due to two-body encounters involving AGB stars of mass
$\sim\,2-8\,$M$_\odot$, whether via dramatic, instantaneous mass loss or via
forming a common envelope system (after a collision) which then
evolves to expel the envelope. This conclusion does not exclude 
the possibility
of collisions involving pre-giant stars. We also note that collisions
involving binaries may be another way to destroy giant or pre-giant
stars (see $\S$9 and Davies \etal 1998).

The broad minima in Figure~\ref{fig:detime} are due to
competition between $\Gamma_1(r)$ and $f_{\rm ce}(r)$. The former is
high at low galactocentric radii and decreases outwards, following the
trends of the
velocity dispersion and density profile. 
$f_{\rm ce}(r)$ is highest when the velocity
dispersion is lowest ($\S$6.2). The outcome is a minimum at about
$r=0.3\,$pc, coincidentally just outside the region of the
observed paucity. 

Another feature is the narrow spread in $\tau_{\rm
ce}$ with impactor mass.  Note that the
curves for the $2\,$M$_\odot$ 
giant encountering $2\,$M$_\odot$, $4\,$M$_\odot$ and $8\,$M$_\odot$
impactors may be underestimating $\tau_{\rm ce}$ somewhat [$\S$6.3]). 
If $\tau_{\rm ce}$ is
independent of $M_2$, then $f_{\rm ce}\propto M_2$ because
$\tau_{\rm ce}\propto 1/(\Gamma_1 f_{\rm ce}) \propto M_2/(\rho f_{\rm
ce})$. Including corrections for uncertainties in $f_{\rm
ce}(r)$ (see $\S$6.3) shows that $\tau_{\rm ce}$ varies by less than a
factor of 2.5 with $M_2$ across a wide range of radii. Clearly 
$\tau_{\rm ce}$ is only weakly dependent on $M_2$.

\section{Further explanations for the observed depletion of giants}

\subsection{Eccentric orbits}

In the calculations of time-scales for two-body collisions
involving giant stars we have assumed that the stars are on circular
orbits, \ie constant galactocentric radii, and so, for a given star, 
the collision rate
is uniform in time. A thermalised stellar system, however, has a
distribution of eccentricities, and so a number of giant stars will be
on orbits that take them very deep into the GC. If the stellar
number-density profile is cusp-like (see $\S$2.4) then the
collision rate in the deep core will be very high, suggesting that
giants on such orbits could be destroyed via collisions when near
their perigees. The time spent by the giants near the perigee is,
however, short, and preliminary investigation suggests that unless the
cusp-like density profile ($\rho \propto r^{-1.8}$) assumed here significantly
underestimates the actual density profile, 
then the number of devastating collisions
involving giants will be insufficient to account for the observed
paucity, even if the impactors are massive black holes rather than
solar-mass stars.

Main-sequence stars as well as giants can be on eccentric orbits that
take them deep into the GC, and so possibly the progenitors of the
giants may be destroyed in collisions. Figure~\ref{fig:timesc} suggests that
collisions involving main-sequence stars could become frequent in the
deep GC if the number-density of stars is cusp-like. Considering
main-sequence stars on eccentric orbits of semi-major axes of about
$0.12\,$pc, the numbers of collisions that they will undergo in the
deep GC is too small to account for the total absence of bright
giants as observed by Genzel \etal (1996). It is worth noting,
however, that the
longevity of low-mass main-sequence stars indigenous to very low
galactocentric radii allows them to undergo
several collisions. 

\subsection{Mass segregation}

\begin{figure}
\psboxto(\hsize;0cm){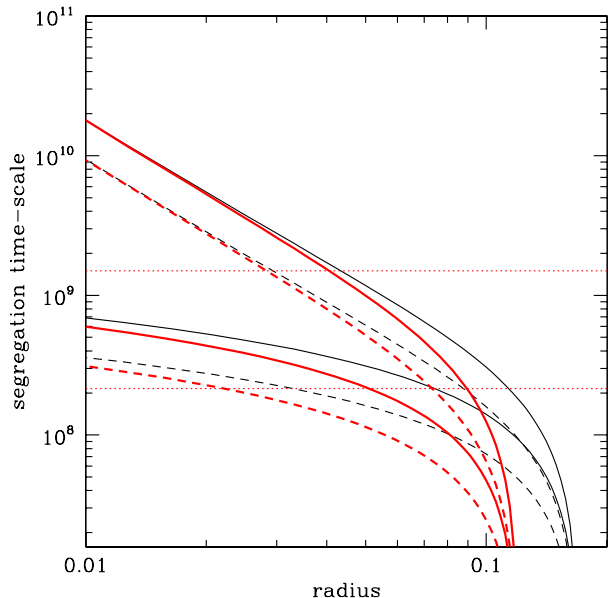}
\caption{Mass segregation time-scales [in years] for stars as a
function of galactocentric radius [parsecs]. 
The time measured is the time taken to sink inwards from an initial radius
of 0.12\,pc (heavy lines) and from 0.17\,pc (light lines) for stars of
mass $2\,$M$_\odot$ (continuous lines) and $4\,$M$_\odot$ (short-dashed
lines). In each case, we have assumed the usual velocity dispersion
profile [Equation~(\protect\ref{eqn:sig1})] and two values of the density
profile: that of Genzel \etal (1996) and the cusp-like profile (see
$\S2.4$); the former gives rise to a longer segregation time-scale and
thus, for a given star, is always the higher line. The
horizontal dotted lines indicate the lifetimes of $2\,$M$_\odot$ stars (upper
line) and $4\,$M$_\odot$ stars (lower line).} 
\label{fig:masseg}
\end{figure}

Stars of mass greater than the
local mean mass will tend to sink deeper into the potential well via
dynamical friction: mass segregation will occur. Mass-segregation
occurs on a time-scale similar to the relaxation time, which for a star of given mass
$M$, is given by:
\begin{equation}
\tau_{\rm rel}= {0.34\sigma(r)^3 \over G^2M\rho(r)ln \Lambda}
\label{eqn:trel}
\end{equation}
(see Binney \& Tremaine 1987), and thus depends on the local values
of density, velocity dispersion and $\Lambda$, the Coulomb potential. 
As $\tau_{\rm rel}$ may
be a function of radius in the GC, so also will be the
mass-segregation rate, which may be estimated as (Hut \etal 1992):
\begin{equation}
{{\rm d}r \over {\rm d}t}\sim -{r\over \tau_{\rm rel}(r)}.
\label{eqn:tseg}
\end{equation}
Thus the time taken for a star's orbit to segregate from radius
$r_{\rm i}$ to $r_{\rm f}$ is
\begin{equation}
\tau_{\rm seg}=-{0.34 \over G^2M}\int^{r_{\rm f}}_{r_{\rm i}} {\sigma^3
\,{\rm d}r \over r\rho ln \Lambda}.
\label{eqn:tauseg}
\end{equation}
For mass segregation to be of interest to the collisional
destruction of stars, we require $\tau_{\rm seg}\leq \tau_{\rm m}$, where
$\tau_{\rm m}$ is the lifetime of a star of mass $M$. If a given star
sinks significantly during its lifetime, then the increase in
collisions that it will undergo in high-density regions could become
important. Disregarding the contribution from $\Lambda$, $\tau_{\rm seg}\propto
1/M$; stars of higher mass sink more quickly than those of lower
mass. However, $\tau_{\rm m}$ decreases strongly with $M$, so
that for $8\,$M$_\odot$ stars sinking from radii of 0.1\,-\,0.2\,pc,
the increase in the numbers of collisions is too small to account for
the absence of the bright giants (Genzel \etal 1996), if such giants
were $8\,$M$_\odot$ stars.
Stars of lower masses, say  $2\,$M$_\odot$ and
$4\,$M$_\odot$, live sufficiently long for significant mass segregation to
occur, as Figure~\ref{fig:masseg} shows. 
The cusp-like density profile gives shorter segregation times than the
profile of Genzel \etal (1996) because the slope of the former tends to
offset the increasing Keplerian ($\sigma\propto r^{-0.5}$) 
velocity dispersion as radius
decreases, whereas the latter profile is
essentially flat at low radius and so the increasing velocity
dispersion actually gives rise to lower sinking rates as radius
decreases.

Although the flatter density profile of Genzel \etal (1996) may allow
$2\,$M$_\odot$ and $4\,$M$_\odot$ stars to sink some distance during
their
lifetimes, the collision rates do not increase strongly enough to make
a difference compared to the case of a star that remains at a constant
galactocentric radius of about 0.12\,pc; this is a  consequence 
of the profile's flatness (Figure~\ref{fig:timesc}). The cusp-like
profile, however, not only allows stars to sink more deeply into the GC but
also provides a strongly increasing collision rate. Assuming the
cusp-like profile, during its lifetime a $4\,$M$_\odot$ star may sink 
down to about 0.02\,pc if it were created at 0.12\,pc. By virtue of their
longevity, stars of mass $2\,$M$_\odot$ may sink to within 0.01\,pc in
the cusp-like profile, even if they were born at 0.17\,pc. Whilst mass
segregating inwards, both stars  will undergo the majority of collisions
whilst on the giant branch, \ie during the latter stages of their lives
after they had had time to sink some distance. A giant star that has
mass segregated from a certain galactocentric radius will have more
collisions than an unsegregated giant star on an eccentric orbit
because the former will spend all its life at low galactocentric radii
whereas the latter is only present in the deep GC when at
periastron. A segregated giant on an eccentric orbit will, of course,
have more collisions than in the other two cases.

Assuming that all other
stars were solar-type, from Equation~(\ref{eqn:zigzag}) we calculate that a
$4\,$M$_\odot$ star indigenous to 0.02\,pc would undergo $\sim80$
collisions, and a $2\,$M$_\odot$ star on a circular orbit at 0.01\,pc
would undergo $\sim400$ collisions. Figure~\ref{fig:detime} shows that
at such low radii, the time-scale for a $2\,$M$_\odot$ giant to
undergo a collision that leads to a common-envelope system is longer
than the stellar lifetime; this is very likely also true for a
$4\,$M$_\odot$ giant. Thus the formation of bound systems cannot
destroy enough of the giants to account for the paucity.

The mass lost from the envelope drops with increasing impact speed, and so
collisions at very low galactocentric radii will involve lower mass
losses than the (small) quantities reported in $\S6$. Individually,
therefore, solar mass impactors cannot destroy the giants, but it may 
be possible that the giants could be
progressively stripped of their envelopes over the many collisions
that they would suffer. Impacts of speeds germane to the very
low-radius GC ($V_\infty \go 800\,$km/s) with solar-mass impactors 
will only remove small amounts of envelope mass from either giant, and
so it is unclear whether solar-mass impactors may be effective in
destroying giants. Impactors of greater mass may be more effective. 
For a target giant in a sea comprised solely
of $8\,$M$_\odot$ impactors (black holes), the number of collisions
will be reduced by a factor of eight compared to solar-type
impactors. Cumulative mass loss via several collisions may remove most
of the giants' envelopes. Alternatively, if a giant is likely to
undergo several collisions with very massive impactors, there may be a
significant probability of a very low impact-parameter collision. This
raises the possibility of the core being pulled out of the envelope by
the massive impactor, thereby destroying the giant. An important
problem with this high-mass impactor scenario is that $2\,$M$_\odot$
stars generally cannot sink as far into the GC if the local mean mass
increases from our assumed $1\,$M$_\odot$ to $8\,$M$_\odot$. 

Collisions involving mass-segregating main-sequence stars 
may be important in the destruction of stars before they evolve into
giants. Unlike the giant phase, the main-sequence stars will undergo
collisions distributed over the range of radii through which the stars
sink. For the main-sequence
counterparts of the two giants considered above, the upper limit of
the numbers collisions suffered is about 30 collisions for the
$2\,$M$_\odot$ star and about two collisions for the $4\,$M$_\odot$ star, for
solar-type impactors and assuming a cusp-like density profile. 
Collisions between main-sequence stars at the
high speeds involved may lead to a significant amount of mass loss and
possibly a merger of the two stars, depending on the impact parameter and
mass ratio of the colliders (Benz \& Hills 1987, 1992; Lai, Rasio \&
Shapiro 1993). It is quite possible that potential giants may be
destroyed via mass loss in collisions whilst on the main-sequence, if
the deep GC density profile is cusp-like. The merging
of the two main-sequence stars into a higher-mass star has the effect
of removing two potential lower-mass giants from the galactic centre
whilst adding a higher-mass star that will eventually evolve into a
massive giant. If such a process occurs, then it is inconsistent with
the paucity observed by Genzel \etal (1996) because massive giants are
extremely luminous whereas the observations show an absence of such
stars in the deep GC. 

Another
consideration must be the process of 
{\it resonant dynamical friction} (see $\S9.3$), which may
take place deep in the GC.  As the
mass-segregation time-scale is similar to the relaxation time-scale, 
resonant dynamical friction may
give rise to segregation time-scales
shorter than the normal segregation time by a factor $\sim 7M_{\rm st}/M_{\rm
bh}$ (see Rauch \& Tremaine 1996), where $M_{\rm st}$ is the enclosed
stellar mass inside a given radius. 
This process should not occur for a system in which the orbits are
isotropic (Rauch \& Tremaine 1996). In a spherically symmetric,
well-mixed distribution of orbits, however, the density profile cannot decrease
inwards. Obviously, the observed paucity of the brightest red giants
means that the density distribution of those stars 
does decrease inwardly within the central $\sim 0.2\,$pc, thereby implying
that the brightest giants' orbits are not well-mixed or
isotropic. However, resonant dynamical friction allows
orbits to drift in angular momentum space, rather than in energy
space and thus their semi-major axes would remain constant.

\subsection{Tidal disruption by a central, massive black hole}

The galactic centre likely harbours a $2\,\times\,10^6\,$M$_\odot$
black hole (Eckart
\& Genzel 1997). Stars will be disrupted if they approach the black
hole to within the tidal radius, which is given by (Hills 1975):
\begin{equation}
R_{\rm t}=\biggl({6M_{\rm bh} \over \pi \rho_1}\biggr)^{1/3}
\label{eqn:rt}
\end{equation}
where $M_{\rm bh}=2\,\times\,10^6\,$M$_\odot$ and $\rho_1$ is the
mean density of the star. We find that $R_{\rm
t}=7.4\,\times\,10^{-4}\,$pc = $3\,\times\,10^5\,$R$_\odot$ for a
star of mass $8\,$M$_\odot$ and radius $243\,$R$_\odot$. For a giant of
mass $2\,$M$_\odot$ and radius $94\,$R$_\odot$, $R_{\rm
t}=4.6\,\times\,10^{-4}\,$pc. In other words, if these giants had
orbital semi-major axes of 0.1\,pc, they would require eccentricities
$> 0.993$ for them to be disrupted. 
Assuming isotropic orbits, the
probability, $P_{\rm t}$, of a star being on an orbit that takes it
inside the tidal radius is given by (\eg Sigurdsson \& Rees 1997):
\begin{equation}
P_{\rm t}\simeq {R_{\rm t} \over d}
\label{eqn:losscone}
\end{equation}
where $d$ is the semi-major axis of the star's orbit. For the star of
mass $8\,$M$_\odot$ and radius $243\,$R$_\odot$, $P_{\rm
t}\sim 7\,\times\,10^{-3}$. 
The observed paucity of giants may be explained by tidal
disruption only if, on time-scales shorter than the
giants' lifetimes, the orbits of {\it all} the giants are scattered
into the phase space occupied by the loss cone. The time-scale for
such scattering to occur is
usually taken to be the relaxation time [Equation~(\ref{eqn:trel})],
but nearly Keplerian potentials (such as that near
the central black hole) can allow faster {\it resonant relaxation} to
occur (Rauch \& Tremaine 1996). This process is important
within the galactocentric radius at which $\mu_{\rm enc}=M_{\rm
st}/(M_{\rm st}+M_{\rm bh})=0.1$ (Rauch \& Ingalls 1998), where $M_{\rm
st}$ is the total enclosed stellar mass. Within the region of
missing bright giants ($\lo 0.2\,$pc), $\mu_{\rm enc}$ 
satisfies this condition,
assuming either the cusp-like density profile or that of Genzel \etal
(1996) (see $\S2.4$). The resonant relaxation time-scale is given by
Rauch \& Tremaine (1996):
\begin{equation}
\tau_{\rm res}={7M_{\rm st} \over M_{\rm bh}} \tau_{\rm rel}.
\label{eqn:tres}
\end{equation}
Well inside the region in which the bright giants are depleted, say at 
0.1\,pc, $\tau_{\rm res}\sim 4\,\times\,10^8$ years, assuming a mean
stellar mass of $\sim 1\,$M$_\odot$ and assuming the density profile of 
Genzel \etal 1996 (the
cusp-like profile actually gives rise to longer resonant relaxation
time-scales as the enclosed stellar mass is higher).
Given the small size of $P_{\rm t}$ and also 
that the resonant relaxation
time is longer than the giant phases of the stars (see
Table~\ref{tbl:time}), we conclude that tidal disruption via
relaxation cannot account for the paucity.

Giants may also enter the loss cone via precession if the GC stellar
distribution (and thus potential) were flattened from spherical, as
the total angular
momentum of a giant in such a potential is not conserved. 
If the flattening is along, say, 
the $z$-direction, then two of the star's angular momentum components ($J_{\rm
x}$ and $J_{\rm y}$) are not conserved whereas $J_{\rm
z}$ is conserved. For giant stars on orbits of $d=0.1\,$pc to pass within
$R_{\rm t}$ of the black hole, the required values of
$J_{\rm z}$ represent only a small fraction of the full range
available. Thus precession into the loss cone in a flattened
asymmetric potential in the GC cannot account for the missing bright giants. 
 
We conclude that tidal disruption cannot account for the paucity of the
bright giant stars.

\subsection{Binary stars}

Interactions between binary stars and red giants in the GC were
considered by Davies \etal (1998). Such interactions
can explain the observed paucity only if the giants are low mass
($2\,$M$_\odot$ rather than $8\,$M$_\odot$) and also if the stars
can be hit at earlier stages of their evolution. A high binary
fraction in the GC population is also required. If these conditions
are satisfied then hard binaries will be more destructive towards
giants than single stars. For stellar collisions to occur in a
binary-single star encounter, 
one expects $R_{\rm min}\lo\,d$. The collision rate for
such encounters is less dependent on the individual size of the
interacting stars than are two-body encounter rates. Thus
main-sequence stars, which are smaller but longer lived than giants,  
will play a greater role in 2+1 body encounters than in single-single
collisions. If a sufficient number of collisions occur in  2+1 body
encounters then the {\it precursors} of the bright-giant stars may be destroyed
before they ascend the giant branch. This will be investigated in
future work.

For sufficiently tight binaries, the size of the primary's Roche lobe limits the
primary's size, or may even lead to the premature end of the primary's
evolution. The binary separation,
$d_{\rm hs}$, for which single cluster members have {\it just} enough kinetic
energy to break up the 
binary is given by:
\begin{equation}
d_{\rm hs}={GM_{\rm a}M_{\rm b}(M_{\rm a}+M_{\rm b}+M_{\rm c}) \over
M_{\rm c}(M_{\rm a}+M_{\rm b})}{1\over \vinf^2}
\label{eqn:dhs}
\end{equation}
(\eg Davies 1995) where $\vinf$ is the relative velocity between the
binary and the single star; $M_{\rm a}$, $M_{\rm b}$ and $M_{\rm c}$ are
respectively the masses
of the two binary components and the mass of a third-body
intruder. Deep in the GC, the mass of the central black hole
dominates, forcing stars to follow Keplerian orbits (see Eckart \& Genzel
1997). As the velocities and velocity dispersions of the stars are
independent of the their masses, the {\it
relative} velocity dispersion between a binary $ab$ and single star $c$ is
therefore 
$\sigma^2_{\rm ab,c}=\sigma^2_{\rm ab}+\sigma^2_{\rm c}=2\sigma^2$
and
the mean-square velocity is ${\langle V_\infty^2\rangle}=3\sigma^2_{\rm
ab,c}=6\sigma^2$. Substituting {$\langle V_\infty^2\rangle$} 
for $\vinf^2$ in Equation~(\ref{eqn:dhs}), 
we find for $\sigma(r)$ appropriate for the GC, $d_{\rm hs}$
is indeed small ($\lo 2
\,$R$_\odot$ for $M_{\rm a}=M_{\rm
b}=M_{\rm c}=1\,$M$_\odot$). 
The time-scale for two objects to pass within, say,  $100\,$R$_\odot$ of each
other deep in the GC is $\sim 10^7\,$years. With the exception of the 
tightest binaries, most binaries  will
have been ionised before the primary could evolve off the main
sequence. We therefore expect most of
the GC giant stars to be single and so the observed paucity of bright
giants cannot be explained in terms of giants' evolution being
restricted in tight binaries.  
Another problem with this scenario is that it does not explain
why the bright giants disappear rather precipitously at about $0.2\,$pc.

\section{Conclusions}

Our chief conclusions may be summarised 
as follows:
\begin{enumerate}
\item The mass losses that result from single-single stellar encounters
involving red giants in the galactic centre are almost always low
($\lo 0.1\,$M$_\odot$). Only a very small number of
collisions will involve sufficient mass loss to destroy the giant. Such
collisions require low $R_{\rm min}$ ($\lo 30\,$R$_\odot$) and
preferentially high-mass impactors ($\go 8\,$M$_\odot$). Such high
mass loss collisions are thus unlikely to account for the
observed paucity of bright red giants. 
\item Mass loss was found to be an increasing function of impactor
mass and decreasing functions of $R_{\rm min}$ and $V_\infty$
reflecting that gravitational forces rather than shock mechanisms
dominated the collisions.
\item Tidal capture of impactors by giant stars is unlikely to occur 
in the galactic centre. Most
collisions involved the impactor passing through the giants and remaining
unbound. This is due to the high velocity dispersion associated with
the galactic centre.
\item The fraction of the impactor's orbital energy ($\Delta E/E$)
deposited in the giant increased with impactor mass and decreased with
increasing $R_{\rm min}$ and $V_\infty$. For sufficiently small $R_{\rm
min}$ and $V_\infty$ the impactor became bound to the giant, 
immediately forming a common envelope system or an eccentric binary system
which would rapidly decay into a common envelope system. The fraction
of collisions resulting in these bound systems was $\lo 0.5$ at a
few pc, decreasing to $\sim\,10^{-5}$ at very low radii, depending
upon the impactor mass. At the radius at which the luminous giants are 
missing ($\sim\,0.2\,$pc), the fraction was $\sim\,10^{-2}\,-\,10^{-3}$.
\item We showed that many of these common envelope systems formed
would lead to total envelope ejection for the $2\,$M$_\odot$ giant and
also for the $8\,$M$_\odot$ giant if the efficiency was high or the
impactor was a high-mass main-sequence star or a stellar
remnant. However, even if all common envelope systems destroy the
giant, the time-scale for collisions which result in these bound
systems was at least the {\it entire} lifetimes of stars of
masses $\go 2\,$M$_\odot$ and so
could not account for the paucity of the giants.
\item The time-scale for creating bound systems was only weakly
dependent of the impactor mass and so can
reveal little about the nature of the undetectable stellar population
of the galactic centre.
\item Giants (or their precursors) within the GC {\it may} be destroyed in
collisions involving high mass losses if their orbits are sufficiently 
eccentric 
to take the stars deeper into the galactic
nucleus where the collision rate is higher. This requires, however,
that we currently significantly underestimate the stellar density deep
in the GC. 
\item Dynamical friction and resonant dynamical friction may
allow stars, originally on wide orbits at about 0.2\,pc (the outer radius
at which bright giants are observed to be missing), to travel deep into
the galactic nucleus, thereby becoming subject to a greatly increased collision
rate. This study does not rule out the possibility of the bright
giants (or their precursors) being destroyed in such collisions. 
This process is only significant, however, if the galactic-centre has
a cusp-like density profile, rather than the observed flattened profile.     
\item The absence of the giant stars is not due to tidal disruption
around a central black hole of mass $2.5\,\times\,10^6\,$M$_\odot$, nor does
it seem likely to be due to single-single stellar 
collisions involving giants. To
explain the paucity in terms of stellar collisions requires
investigation into the collisional effects of binary stars upon
pre-giant stars. Hard binaries in the galactic centre may well have
separations too small for the primary to evolve into a giant. This
also, however, means that most binaries will have been ionised and so
most giants will be single stars (unless, perhaps, they exchange into
a binary). It also does not
seem to explain why the numbers of bright giants plummet rapidly at
$\sim\,0.2\,$pc.  
\end{enumerate}

\section*{ACKNOWLEDGEMENTS}

	We thank J. Brinchmann, P. Eggleton, J. Hurley, J. Magorrian,
 S. Sigurdsson and C. Tout, of the IoA and R. Genzel (MPE-Garching) 
for valuable
discussions. We also thank the referee for insightful comments. 
Simulations were 
performed at T6 (Los Alamos National laboratory) and at the IoA; we
are grateful to M. Warren \& M.P. Goda (T6, LANL) for technical support.  
This work was supported through an IGPP research grant
at Los Alamos. VCB was supported by a PPARC grant. 
MBD gratefully acknowledges the support of a URF from the
Royal Society.


\begin{thebibliography}{}
\newcommand{\jn}{\textbf}


\bibitem[]{} Binney J.J., Tremaine S., 1987, Galactic Dynamics.
Princeton
\bibitem[]{} Bahcall J.N., Wolf R.A., 1976, ApJ, 209, 214
\bibitem[]{} Benz W., 1990, The Numerical Modelling of Nonlinear
Stellar Pulsations, J.R. Buchler (ed.), 269
\bibitem[]{} Benz W., Hills J.G., 1987, ApJ, 323, 614
\bibitem[]{} Benz W., Hills J.G., 1992, ApJ, 389, 546
\bibitem[]{} Blum R.D., Sellgren K., DePoy D.L., 1996, ApJ, 470, 864
\bibitem[]{} Davies M.B., 1995, MNRAS, 276, 887
\bibitem[]{} Davies M.B., Benz W., Hills J.G., 1991, ApJ, 381, 449
\bibitem[]{} Davies M.B., Benz W., Hills J.G., 1992, ApJ, 401, 246
\bibitem[]{} Davies M.B., Blackwell R., Bailey V.C., Sigurdsson S.,
1998, MNRAS, 301, 745
\bibitem[]{} de Kool M., 1990, ApJ, 358, 189
\bibitem[]{} Eckart A., Genzel R., 1997, MNRAS, 284, 576
\bibitem[]{} Eckart A., Genzel R., Hofmann R., Sams B.J.,
Tacconi-Garman L.E., 1995, ApJ, 445, L23
\bibitem[]{} Eggleton P.P., 1983, ApJ, 268, 368
\bibitem[]{} Genzel R., Hollenbach D., Townes C.H., 1994,
Rep. Prog. Phys. 57, 417
\bibitem[]{} Genzel R., Thatte N., Krabbe A., Kroker H.,
Tacconi-Garman L.E., 1996, ApJ, 472, 153 
\bibitem[]{} Han Z., Podsiadlowski P., Eggleton P.P., 1994, MNRAS,
270, 121
\bibitem{} Hills J.G., 1975, Nat, 254, 295
\bibitem{} Hut P., McMillan S., Romani R., 1992, ApJ, 389, 527
\bibitem{} Johnson H.L., 1966 ARA\&A, 4, 193
\bibitem{} Morris M., 1993, ApJ, 408, 496
\bibitem[]{} Pols O.R., Schroder K.-P., Hurley J.R., Tout C.A.,
Eggleton P.P., 1999, in prep
\bibitem[]{} Pols O.R., Tout C.A., Eggleton P.P., Han Z., 1995,
MNRAS, 274, 964
\bibitem[]{} Rauch K.P., Ingals B., 1998, MNRAS, 299, 1231
\bibitem[]{} Rauch K.P., Tremaine S., 1996, New Astron., 1, 149 (RT96)
\bibitem[]{} Quinlan G.D., Hernquist L., Sigurdsson S., 1995, ApJ,
440, 554
\bibitem[]{} Rasio F.A., Shapiro S.L., 1991, ApJ, 377, 559
\bibitem[]{} Rieke G.H., Rieke M.J., Paul A.E., 1989, ApJ, 336, 623
\bibitem[]{} Saha P., Biknell V., McGregor P.J., 1996, ApJ, 467, 636
\bibitem[]{} Sandquist E.L., Taam R.E., Chen X., Bodenheimer P.,
Burkert A., 1998, ApJ, 500, 909
\bibitem[]{} Sigurdsson S., Rees M., 1997, MNRAS, 284, 318
\bibitem[]{} Taam R.E., Bodenheimer P., 1989, ApJ, 337, 849  
\bibitem[]{} Tout C., Pols O.R., Eggleton P.P., Han Z., 1996, MNRAS,
281, 257

\end{thebibliography}
\end{document}